\documentclass[10pt,twocolumn,letterpaper]{article}

\usepackage{iccv}
\usepackage{times}
\usepackage{epsfig}
\usepackage{graphicx}
\usepackage{amsmath}
\usepackage{amssymb}
\usepackage{mathtools}
\usepackage{amssymb,amsmath,amsthm}
\usepackage{flushend}
\usepackage{subcaption}

\graphicspath{
    {./Figs/}
    }

\newtheorem{theorem}{Theorem}

\usepackage[pagebackref=true,breaklinks=true,letterpaper=true,colorlinks,bookmarks=false]{hyperref}

\iccvfinalcopy 

\def\iccvPaperID{5880} 

\newcommand{\bPhi}{\mathbf{\Phi}}
\newcommand{\C}{\mathbf{C}}
\newcommand{\D}{\mathbf{D}}
\newcommand{\K}{\mathbf{K}}
\newcommand{\ba}{\mathbf{a}}
\newcommand{\bb}{\mathbf{b}}
\newcommand{\bLambda}{\Lambda}

\ificcvfinal\fi
\begin{document}

\title{OperatorNet: Recovering 3D Shapes From Difference Operators}

\author{Ruqi Huang\thanks{denotes equal contribution.}\\
LIX, Ecole Polytechnique\\
{\tt\small rqhuang88@gmail.com}
\and
Marie-Julie Rakotosaona\footnotemark[1]\\
LIX, Ecole Polytechnique\\
{\tt\small mrakotos@lix.polytechnique.fr}
\and 
Panos Achlioptas\\
Stanford University\\
{\tt\small optas@cs.stanford.edu}
\and 
Leonidas Guibas\\
Stanford University\\
{\tt\small guibas@cs.stanford.edu}
\and
Maks Ovsjanikov\\
LIX, Ecole Polytechnique\\
{\tt\small maks@lix.polytechnique.fr}
}

\maketitle

\begin{abstract}
  This paper proposes a learning-based framework for reconstructing 3D
  shapes from functional operators, compactly encoded as small-sized
  matrices. To this end we introduce a novel neural architecture, called
  \emph{OperatorNet}, which takes as input a set of linear operators
  representing a shape and produces its 3D embedding. We demonstrate
  that this approach significantly outperforms previous purely
  geometric methods for the same problem. Furthermore, we introduce a novel
  functional operator, which encodes the extrinsic or pose-dependent
  shape information, and thus complements purely intrinsic
  pose-oblivious operators, such as the classical Laplacian. Coupled
  with this novel operator, our reconstruction network achieves very
  high reconstruction accuracy, even in the presence of incomplete
  information about a shape, given a soft or functional
  map expressed in a reduced basis. Finally, we demonstrate that the multiplicative
  \emph{functional algebra} enjoyed by these operators can be used to
  synthesize entirely new unseen shapes, in the context of shape
  interpolation and shape analogy applications.

\end{abstract}
\vspace{-3mm}

\section{Introduction}\label{sec:intro}

Encoding and reconstructing 3D shapes is a fundamental problem in Computer Graphics, Computer Vision and related fields. Unlike images, which enjoy a canonical representation, 3D shapes are encoded through a large variety of representations, such as point clouds, triangle meshes and volumetric data, to name a few. Perhaps even more importantly, 3D shapes may undergo a diverse set of transformations, ranging from rigid motions to complex non-rigid and articulated deformations, that impact these representations.

The representation issues have become even more prominent with the recent advent of learning-based techniques, leading to a number of solutions for learning directly on geometric 3D data \cite{bronstein2017geometric}. This is challenging, as point clouds and meshes lack the regular grid structure exploited by convolutional architectures. In particular, devising representations that are well-adapted for both shape analysis and especially \emph{shape synthesis} remains difficult. For example, several methods for shape interpolation have been proposed by designing deep neural networks, including auto-encoder architectures, and interpolating the latent vectors learned by such networks~\cite{Surfnet, achlioptas2017learning} . Unfortunately, it is not clear if the latent vectors lie in a linear vector space, and thus linear interpolation can lead to unrealistic intermediate shapes.

\begin{figure}[t!]
  \centering
  \includegraphics[width=0.95\linewidth]{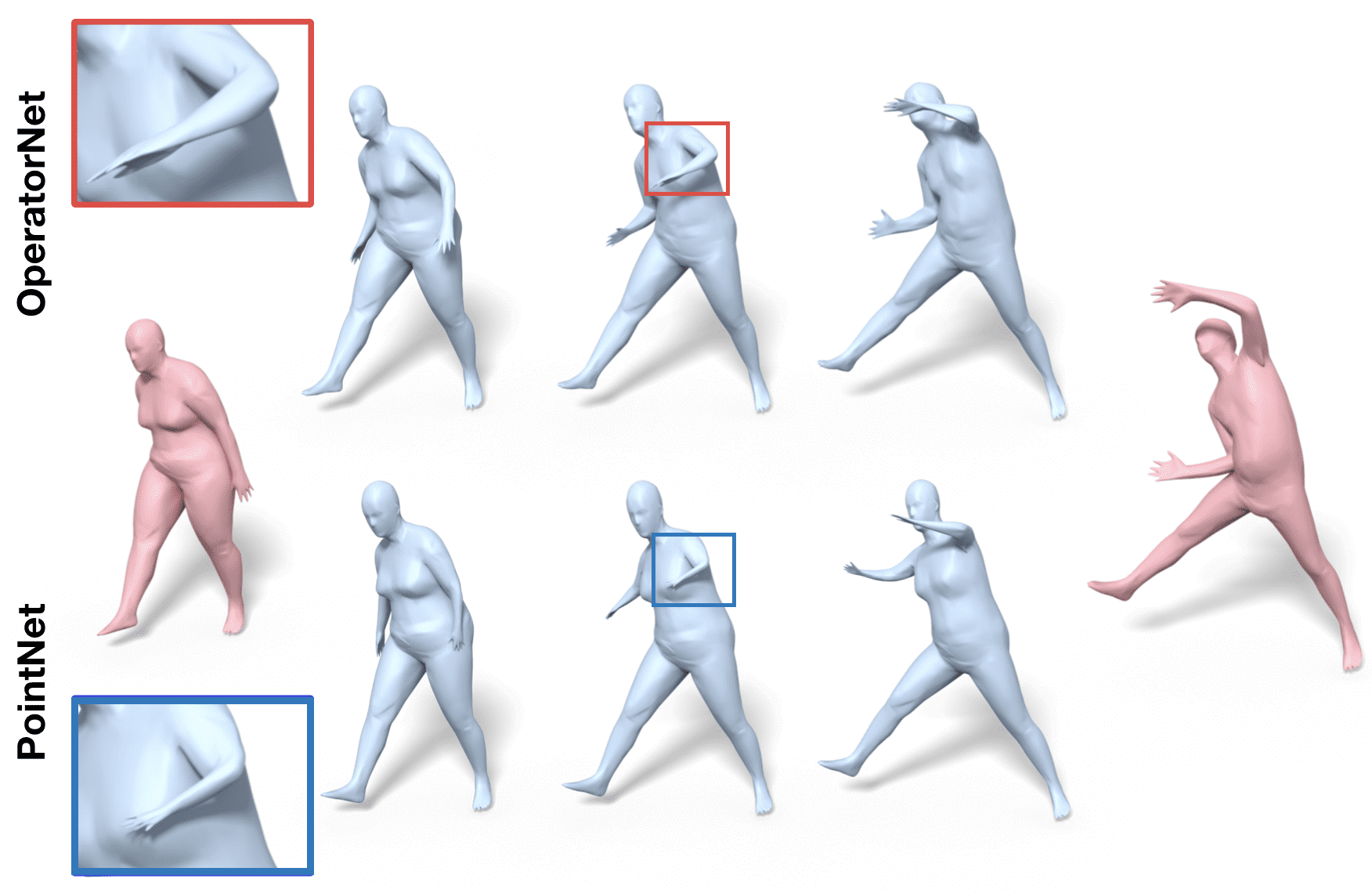}
\caption{\label{fig:teaser_interpolation} Shape interpolation via OperatorNet (top) and PointNet autoencoder (bottom). Our interpolations are more smooth and less distorted.}
\vspace{-2em}
\end{figure}

In this paper, we show that 3D shapes can not only be compactly encoded as linear functional operators, using the previously proposed shape difference operators~\cite{Rustamov2013}, but that this representation lends itself very naturally to learning, and allows us to \emph{recover} the 3D shape information, using a novel neural network architecture which we call OperatorNet. Our key observations are twofold: first we show that since shape difference operators can be stored as canonical matrices, for a given choice of basis, they enable the use of a convolutional neural network architecture for shape recovery. Second, we demonstrate that the \emph{functional algebra} that is naturally available on these operators can be used to synthesize new shapes, in the context of shape interpolation and shape analogy applications. We argue that because this algebra is well-justified theoretically, it also leads to more accurate results in practice, compared to commonly used linear interpolation in the latent space (see Figure~\ref{fig:teaser_interpolation}).  

The shape difference operators introduced in~\cite{Rustamov2013}, have proved to be a powerful tool in shape analysis, by allowing to characterize each shape in a collection as the ``difference'' to some base geometry. These difference operators encode precise information about how and where each shape differs from the base, but also, due to their compact representation as small matrices, enable efficient exploration of \emph{global} variability within the collection. Inspired by the former perspectives, purely geometric approaches~\cite{boscaini2015shape, Corman2017} have been proposed for shape reconstruction from shape differences. Though theoretically well-justified, these approaches rely on solving difficult non-linear optimization problems and require strong regularization for accurate results, especially when truncated bases are used.

Our OperatorNet, on the other hand, leverages the information encoded at both the pairwise level and the collection level by using the shape collection to guide the reconstruction. It is well-known that related shapes in a collection often concentrate near a low-dimensional manifold in shape space~\cite{Schulz2017, SMPL}. In light of this, the shape difference operators can help to both encode the geometry of the individual shapes, but also help to learn the constrained space of realistic shapes, which is typically ignored by purely geometric approaches. Finally, they also allow to encode differences between shapes with different discretizations by relying on \emph{functional} maps, rather than, e.g., pointwise bijections.

In addition to demonstrating the representative power of the shape differences in a learning framework, we also extend the original formulation in~\cite{Rustamov2013}, which only involves intrinsic (i.e., invariant to isometric transformations) shape differences, with a novel \emph{extrinsic} difference operator that facilitates pose-dependent embedding recovery. Our formulation is both simpler and robuster compared to previous approaches, e.g.~\cite{Corman2017}, and, as we show below, can more naturally be integrated in a unified learning framework.

To summarize, our contributions are as follows:

\begin{itemize}
\setlength\itemsep{-0.5em}
    \item We propose a learning-based pipeline to reconstruct 3D shapes from a set of difference operators. 
    \item We propose a novel formulation of extrinsic shape difference, which complements the intrinsic operators formulated in~\cite{Rustamov2013}.
    \item We demonstrate that by applying algebraic operations on shape differences, we can synthesize new operators and thus new shapes via OperatorNet, enabling shape manipulations such as interpolation and analogy. 
\end{itemize}
\section{Related Work}\label{sec:related}

\paragraph{Shape Reconstruction}
Our work is closely related to shape reconstruction from intrinsic operators, which was recently considered in
\cite{boscaini2015shape,Corman2017} where several advanced, purely geometric optimization techniques have been proposed
that give satisfactory results in the presence of full information \cite{boscaini2015shape} or under strong 
regularization \cite{Corman2017}. These works have also laid the theoretical foundation for shape recovery by
demonstrating that shape difference operators, in principle, contain complete information necessary for recovering the
shape embedding (e.g. Propositions 2 and 4 in \cite{Corman2017}). On the other hand, these methods also
highlight the practical challenges of reconstructing a shape without any
knowledge of the collection or ``shape space'' that it belongs to. In contrast, we show that by leveraging such
information via a learning-based approach, realistic 3D shapes can be recovered efficiently from their shape difference
representation, and moreover that entirely new shapes can be synthesized using the algebraic structure of difference
operators, e.g., for shape interpolation.
\vspace{-2mm}
\paragraph{Shape Representations for Learning.}

Our work is related to the recent techniques aimed at applying deep learning methods to
shape analysis. One of the main challenges is defining a meaningful notion on convolution, while
ensuring invariance to basic transformations, such as rotations and translations. Several techniques have
been proposed based on e.g., Geometry Images \cite{sinha2016deep}, volumetric
\cite{maturana2015voxnet,wang2017cnn}, point-based \cite{qi2016_pointnet} and multi-view approaches
\cite{qi2016volumetric}, as well as, very recently intrinsic techniques that adapt convolution to
curved surfaces \cite{masci2015geodesic,boscaini2016learning,poulenard2018multi} (see also
\cite{bronstein2017geometric} for an  overview), and even via toric covers
\cite{maron2017convolutional}, among many others.

Despite this tremendous progress in the last few years, defining a shape representation that is compact, lends itself
naturally to learning, while being invariant to the desired class of transformations (e.g., rigid motions) and not
limited to a particular topology, remains a challenge. As we show below, our representation is well-suited for learning
applications, and especially for encoding and recovering geometric structure information. 
We note that a recent work that is closely related to ours is the characteristic shape differences proposed in~\cite{limitshape}. That work is primarily focused on \emph{analyzing} shape collections, rather than on shape \emph{synthesis} that we target.
\vspace{-2mm}
\paragraph{Shape Space} 
Exploring the structure of shape spaces has a long and extensive research history.  Classical PCA-based models,
e.g. ~\cite{scape,hasler2009statistical}, and more recent shape space models, adapted to specific shape classes such as
humans~\cite{SMPL} or animals~\cite{smal}, or parametric model collections~\cite{Schulz2017}, all typically leverage
the fact that the space of ``realistic'' shapes is significantly smaller than the space of all possible embeddings. This
has also recently been exploited in the context of learning-based shape synthesis applications for shape completion
\cite{litany2018deformable}, interpolation \cite{ben2018multi} and point cloud reconstruction
\cite{achlioptas2017learning} among others. These techniques heavily leverage the recent proliferation of large data
collections such as DFAUST~\cite{dfaust} and Shapenet~\cite{shapenet2015} to name a few. At the same time, it is not
clear if, for example, the commonly used \emph{linear interpolation} of latent vectors is well-justified, leading to
unrealistic synthesized shapes. Instead, the shape difference operators that we use satisfy a well-founded
multiplicative algebra, which, as we show below, can be used to create realistic synthetic shapes.

\section{Preliminaries and Notations}\label{sec:background}

\paragraph{Discretization of Shapes} Throughout this paper, we assume that a shape is given as a triangle mesh
$(\mathcal{V}, \mathcal{F})$, where $\mathcal{V} = \{v_1, v_2, \cdots, v_n\}$ is the vertex set, and
$\mathcal{F} = \{(v_i, v_j, v_k)| v_i, v_j, v_k\in \mathcal{V}\}$ is the set of faces encoding the connectivity
information.

\paragraph{Laplace-Beltrami Operator} To each shape $S$, we associate a discretized Laplace-Beltrami operator,
$\mathcal{L}\coloneqq A^{-1} W$, using the standard cotangent weight scheme ~\cite{meyer03, pinkall1993computing}, where $W$
is the cotangent weight (stiffness) matrix, and $A$ is the diagonal lumped area (mass) matrix.  Furthermore, we denote
by $\mathbf{\Lambda}, \bPhi$, respectively the diagonal matrix containing the $k$ smallest eigenvalues and the
corresponding eigenvectors of $S$, such that $W\mathbf{\Phi} = A \mathbf{\Phi \Lambda}$.  In particular, the eigenvalues
stored in $\mathbf{\Lambda}$ are non-negative and can be ordered as $0 = \lambda_1 \leq \lambda_2 \leq \cdots$. The columns of
$\mathbf{\Phi}$ are sorted accordingly, and are orthonormal with respect to the area matrix, i.e., $\mathbf{\Phi}^T A
\mathbf{\Phi} = \textbf{I}_{k \times k}$, the
$k \times k$ identity matrix.  It is well-known that Laplace-Beltrami eigenbasis provides a multi-scale encoding
of a shape~\cite{levy2006}, and allows to approximate the space of functions via a subspace spanned by
the first few eigenvectors of $\bPhi$.

\vspace{-2mm}
\paragraph{Functional Maps} The functional map framework was introduced in~\cite{Functional} primarily as an alternative
representation of maps across shapes. In our context, given two shapes $S_0, S_1$ and a point-wise map $T$ from $S_1$ to
$S_0$, we can express the functional map $\mathbf{C}_{01}$ from $S_0$ to $S_1$, as follows:
\begin{equation}
    \C_{01} = \bPhi_1^T A_1 \Pi_{01} \bPhi_0 \label{equ:fmap} .
\end{equation}
Here, $A_1$ is the area matrix of $S_1$, and $\Pi_{01}$ is a binary matrix satisfying $\Pi_{01}(p, q) = 1$ if $T(p) = q$
and $0$ otherwise. Note that $\C_{01}$ is a $k_1 \times k_0$ matrix, where $k_1, k_0$ is the number of basis
functions chosen on $S_1$ and $S_0$. This matrix allows to transport functions as follows: if $f$
is a function on $S_0$ expressed as a vector of coefficients $\mathbf{a}$, s.t. $f = \bPhi_0 \mathbf{a}$, then $\C_{01} \mathbf{a}$
is the vector of coefficients of the corresponding function on $S_1$, expressed in the basis of $\bPhi_1$.

In general, not every functional map matrix arises from a point-wise
map, and the former might include, for
example, soft correspondences, which map a point to a probability density function. All of the tools that we develop below can
accommodate such general maps. This is a key advantage of our approach, as it does not rely on all shapes having the same
number of points, and only requires the knowledge of functional map matrices,
which can be computed using existing techniques \cite{ovsjanikov2017computing,litany2017deep}.

\vspace{-2mm}
\paragraph{Intrinsic Shape Difference Operators}
Finally, to represent shapes themselves, we use the notion of shape difference operators proposed
in~\cite{Rustamov2013}. Within our setting, they can be summarized as follows: given a base shape $S_0$, an arbitrary
shape $S_i$ and a functional map $\C_{0i}$ between them, let $\K_0$ (resp. $\K_i$) be a positive semi-definite matrix,
which defines some inner product for functions on $S_0$ (resp. $S_i$) expressed in the corresponding bases. Thus, for a
pair of functions $f, g$ on $S_0$ expressed as vectors of coefficients $\ba, \bb$, we have $<f, g> = \ba^T \K_0 \bb.$

Note that these two inner products $\K_0, \K_i$ are not comparable, since they are expressed in different
bases. Fortunately, the functional map $\C_{0i}$ plays a role of basis synchronizer. Thus, a shape difference operator,
which captures the difference between $S_0$ and $S_i$ is given simply as:
\begin{equation}
	\D^K_{0i} = \K_0^{+} ( \C_{0i}^T \K_i \C_{0i}), \label{equ:shapediff}
\end{equation}
where $^{+}$ is the Moore-Penrose pseudo-inverse.

The original work \cite{Rustamov2013} considered two intrinsic inner products, which using the notation above, can be
expressed as: $\K^{L^2} = \mathbf{Id}$, and $\K^{H^1} = \mathbf{\Lambda}$.

These inner products, in turn lead to the following shape differences operators: 
\begin{align}
    &\mbox{Area-based ($L^2$): } & \D_{0i}^A =  & \C_{0i}^T \C_{0i}, \label{equ:area-SD}\\ 
    &\mbox{Conformal ($H^1$): } & \D_{0i}^C = & \bLambda_0^{+} \C_{0i}^T \bLambda_i \C_{0i}, \label{equ:conf-SD}
\end{align}

These shape difference operators have several key properties. First,
they allow to represent an arbitrary shape $S_i$, as a pair of
matrices of size $k_0 \times k_0$, independent of the number of
points, by requiring only a functional map between the base shape
$S_0$ and $S_i$. Thus, the size of this representation can be
controlled by choosing an appropriate value of $k_0$ which allows to
gain multi-scale information about the geometry of $S_i$, from the
point of view of $S_0$. Second, and perhaps more importantly, these
matrices are invariant to rigid (and indeed any intrinsic isometry)
transformation of $S_0$ or $S_i$. Finally, previous works
\cite{Corman2017} have shown that shape differences \emph{in
  principle} contain complete information about the intrinsic geometry
of a shape. As we show below these properties naturally enable the use
of learning applications for shape recovery.

\paragraph{Functoriality of Shape Differences}
Another useful property of the shape difference operators is
\emph{functoriality}, shown in \cite{Rustamov2013}, and which we
exploit in our shape synthesis applications in Section~\ref{sec:app}.
Given shape differences $\D_{0i}, \D_{0j}$ of shapes $S_i$ and $S_j$
with respect to a base shape $S_0$, functoriality allows to compute
the difference $\D_{ij}$, without functional maps between $S_i$ and
$S_j$. Namely (see Prop. 4.2.4 in \cite{corman2016}):
\begin{equation}\label{equ:functoriality}
	\D_{ij} = \C_{0i} \D_{0i}^{+} \D_{0j}  \C_{0i}^{-1}
\end{equation} 
Intuitively, this means that shape differences naturally satisfy the
\emph{multiplicative algebra}: $\D_{0i}\D_{ij} = \D_{0j}$, up to a change of
basis ensured by $\C_{0i}$.

\begin{figure}[t!]
  \centering
  \includegraphics[width=\linewidth]{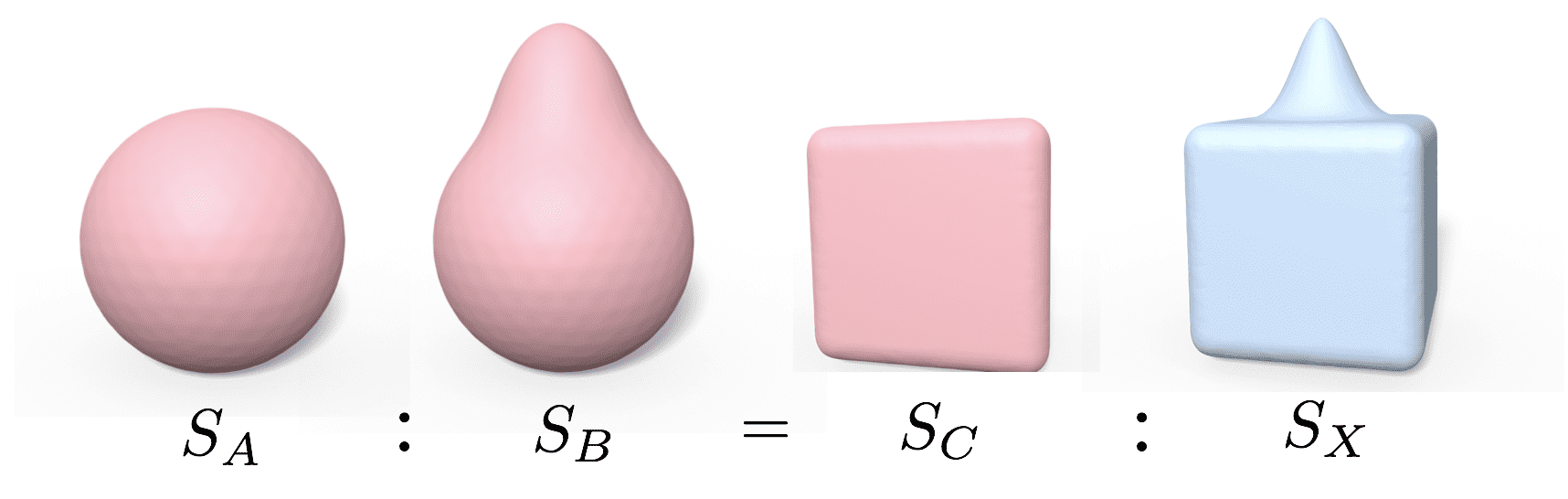}
\caption{\label{fig:analogy_illustration} Illustration of shape analogy. }
\vspace{-1em}
\end{figure}

This property can be used for \emph{shape analogies}: given shapes
$S_A, S_B$ and $S_C$, find $S_X$ such that $S_X$ relates to
$S_C$ in the same way as $S_B$ relates to $S_A$ (see the illustration in Figure~\ref{fig:analogy_illustration}). This can be solved
by looking for a shape $X$ that satisfies: $\C_{0C}^{+} \D_{CX} \C_{0C} = \C_{0A}^{+} \D_{AB}
\C_{0A}$. In our application, we first create an
appropriate $\D_{0X}$ and then use our network to synthesize the
corresponding shape.

Finally, the multiplicative property also suggests a way of
interpolation in the space of shape differences. Namely, rather than using basic
linear interpolation between $\D_{0i}$ and $\D_{0j}$, we interpolate
on the Lie algebra of the Lie group of shape differences, using the
exponential map and its inverse, which leads to:
\begin{equation}\label{equ:intpolation}
	\D(t)  = \exp((1-t)\log(\D_{0i}) + t\log(\D_{0j})), t \in [0, 1].
\end{equation}
Here $\exp$ and $\log$ are matrix exponential and logarithm
respectively. Note that, around identity, the linearization provided
by the Lie algebra is exact, and we have observed it to produce very
accurate results in general.

\section{Extrinsic Shape Difference}\label{sec:extdiff}

In our (discrete) setting, with purely \emph{intrinsic} information one at the best can determine the edge lengths of the mesh. Recovering the shape from its edge lengths, while possible in certain simple scenarios, nevertheless often leads to ambiguities, as highlighted in \cite{Corman2017}. 
To alleviate such ambiguities, we propose to augment the existing intrinsic shape differences with a novel \emph{extrinsic} shape difference operator, and in turn boosts our reconstruction.

One basic approach to combine extrinsic information with the multi-scale Laplace-Beltrami basis is to project the 3D coordinate functions onto the basis, to obtain three vectors of coefficients (one for each $x, y, z$ coordinates): $\mathbf{f}  = \bPhi^{+} X$, where $X$ is the $n_{V} \times 3$ matrix of vertex coordinates \cite{levy2006,kovnatsky2013coupled}. Unfortunately representing a shape through $\mathbf{f}$, though being multi-scale and compact, is not rotationally invariant, and does not provide information about intrinsic geometry. For example, interpolation of coordinate vectors can easily lead to loss of shape area.

Another option, which is more compatible with our approach and is rotationally invariant, is to encode the inner products of coordinate functions on each shape using the Gram matrix $G = X X^T$. Expressing $G$ in the corresponding basis, and using Eq.~\eqref{equ:shapediff} gives rise to a shape difference-like representation of the coordinates. Indeed, the following theorem (see proof in Appendix~\ref{sec:proof}) guarantees that the resulting representation contains the same information, up to rotational invariance, as simply projecting the coordinates onto the basis.

\begin{theorem}\label{thm:recovery}
Let $\mathbf{G} = \bPhi^T A X X^T A \bPhi$ be the extrinsic inner product encoded in $\Phi$, then one can recover the projections of the coordinate functions, $X$, on the subspace spanned by $\bPhi$ from $\mathbf{G}$, up to a rigid transformation. 
In particular, when $\bPhi$ is a complete full basis, the recovery of $X$ is exact. 
\end{theorem}

\begin{figure}[t!]
  \centering
  \includegraphics[width=0.95\linewidth]{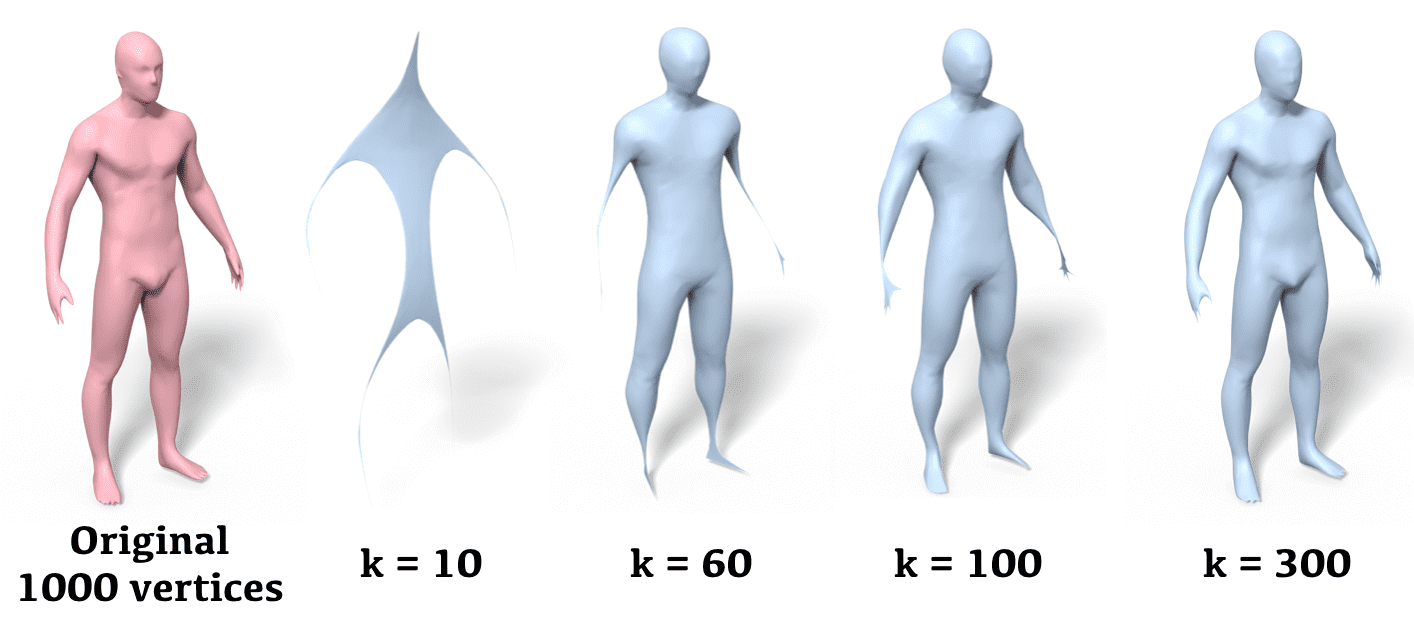}
\caption{\label{fig:recovery} From left to right: original shape with 1000 vertices, the recovered embedding from $\mathbf{G}$ encoded in the leading k = 10, 60, 100 and 300 eigenbasis of the original shape.}
\vspace{-2mm}
\end{figure}

As an illustration of Theorem~\ref{thm:recovery}, we show in Figure~\ref{fig:recovery} the embeddings recovered from $\mathbf{G}$ when the number of basis functions in $\bPhi$ ranges from 10 to 300.

However, the rank of the Gram matrix $\mathbf{G}$ of a shape is at most $3$, meaning that the majority of its eigenvalues are zero. This turns out to be an issue in applications, where gaining information about the \emph{local} geometry of the shape is important, for example in our shape analogies experiments.

To compensate for this rank deficiency, we make the extrinsic inner product Laplacian-like: 
\begin{equation}
    {E}^D(i, j) = \begin{cases}
    -E(i, j) &\text{if $i\neq j$,}\\
    \sum_{i\neq j} E(i, j) &\text{i = j.}
    \end{cases}\label{equ:ext-innerproduct}
\end{equation}
  Where $E(i,j)$ is $\Vert v_i - v_j \Vert^2 A(i, i) A(j, j)$, i.e., the squared Euclidean distance between points $v_i, v_j$ on the shape, weighted by the respective vertex area measures. 
Since $E^D$ can be regarded as the Laplacian of a complete graph, all but one of its eigenvalues are strictly positive.

It is worth noting that the Gram matrix and the squared Euclidean distance matrix are closely related and can be recovered from each other as is commonly done in the Multi-Dimensional Scaling literature~\cite{MDS}.

To summarize, given a base shape $S_0$, another shape $S_i$ and a functional map $\C_{0i}$ we encode the extrinsic information of $S_i$ from the point of view of $S_0$ as follows:
\begin{equation} \label{eq:interp}
    \D^E_{i} = (\bPhi_0^T {E}_0^D \bPhi_0)^{+}(\C_{0i}^T \bPhi_i^T {E}_i^D \bPhi_i \C_{0i}). 
\end{equation}

\begin{figure}[t!]
  \centering
  \includegraphics[width=0.95\linewidth]{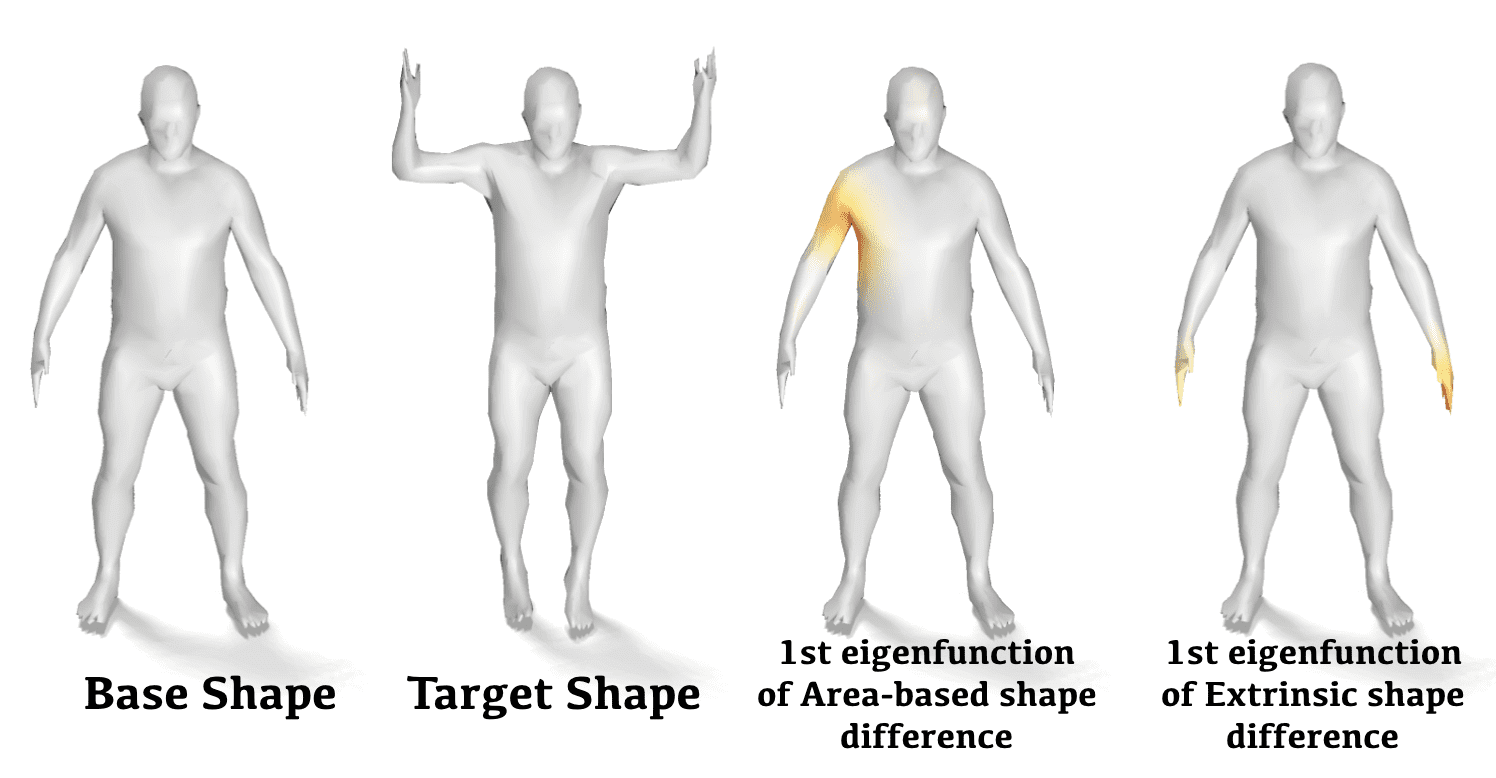}
\caption{\label{fig:eigenfuncs} A pair of shapes are compared. The most area (resp. extrinsic) distorted region is captured by the leading eigenfunction of the area-based (resp. extrinsic) shape difference. }
\vspace{-2mm}
\end{figure}

In Figure~\ref{fig:eigenfuncs}, we compute $\D^A$ and $\D^E$ of the target shape with respect to the base, and color code their respective eigenfunctions associated with the largest eigenvalue on the shapes to the right. As argued in \cite{Rustamov2013} these functions capture the areas of highest distortion between the shapes, with respect to the corresponding inner products. Note that the eigenfunction of $\D^A$ captures the armpit where the local area shrinks significantly, while that of $\D^E$ captures the hand, where the \emph{pose changes} are evident.

Note that in~\cite{Corman2017}, the authors also propose a shape difference formulation for encoding extrinsic information, which is defined on the shape offset using the surface normal information. 
However, their construction can lead to instabilities, and moreover, it only gives information about \emph{local} distances, making it hard to recover large changes in pose.

\section{Network Details}\label{sec:network}

\paragraph{Problem Setup}
Our general goal is to develop a neural network capable of recovering the coordinates of a shape, given its representation as a set of shape difference matrices. We therefore aim to solve the same problem considered in \cite{boscaini2015shape,Corman2017}. However, unlike these purely geometric methods, we also leverage a collection of training shapes to learn and constrain the reconstruction to the space of realistic shapes.

Thus, we assume that we are given a collection of shapes, each represented by a set of shape difference operators with respect to a fixed base shape. We also assume the presence of a point-wise map from the base shape to each of the shapes in the collection, which allows us to compute the ``ground truth'' embedding of each shape. We represent this embedding as three coordinate functions on the shape. Our goal then is to design a network, capable of converting the input shape difference operators to the ground truth coordinate functions.

At test time, we use this network to reconstruct a target shape given only the shape difference operators with respect to the base shape. Importantly, these shape difference operators only require the knowledge of a \emph{functional map} from the base shape, and can thus arise from shapes with different discretizations, or can be synthesized directly for shape analogies or interpolations applications.

\begin{figure}[t!]
  \centering
  \includegraphics[width=0.9\linewidth]{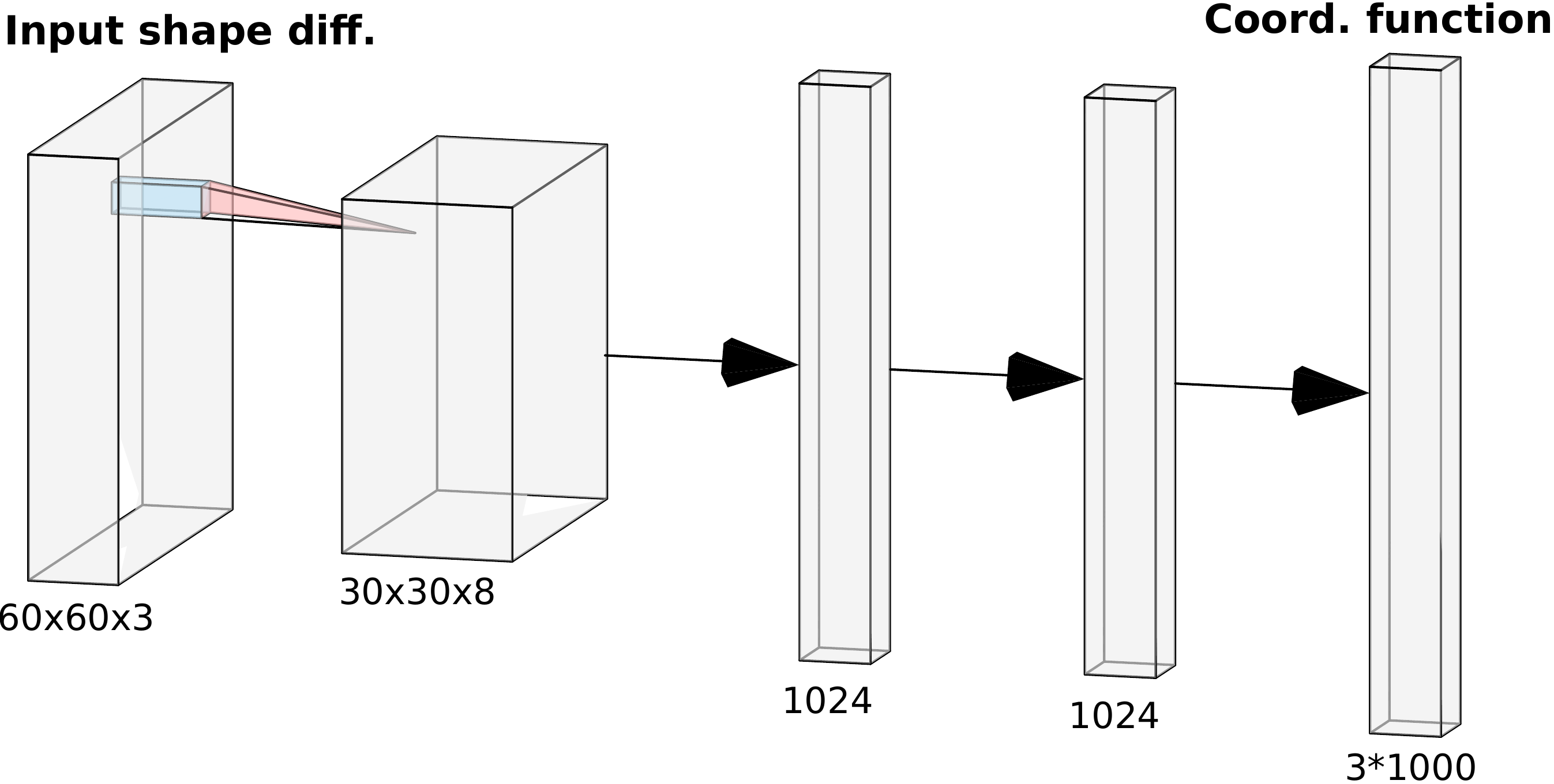}
\caption{\label{fig:architecture} OperatorNet architecture. The inputs of the network are shape difference matrices considered as channels. It outputs the coordinate functions of the shape. The first part (left) of the network consists of a convolutional encoder while the second part (right) is a fully-connected decoder built with dense layers.}
 \vspace{-2mm}
\end{figure}

\vspace{-2mm}
\paragraph{Architecture}
To solve the problem above we developed the OperatorNet architecture, which takes as input shape difference matrices and outputs coordinate functions. Our network has two modules: a shallow convolutional encoder and a 3 layer dense decoder as shown in Figure~\ref{fig:architecture}.

The grid structure of shape differences is exploited by the encoder through the use of convolutions. Note however that translation invariance does not apply to these matrices. 

After comparing multiple depths of encoders, we select a shallow version as it performs the best in practice, implying that the shape difference representation already encodes meaningful information efficiently. Moreover, as shown in \cite{Corman2017} the edge lengths of a mesh can be recovered from intrinsic shape differences through a series of least squares problems, hinting that increasing the depth of the network and thus the non-linearity might not be necessary with shape differences.   

On the other hand, the decoder is selected for its ability to transform the latent representation to coordinate functions for reconstruction and synthesis tasks.    
\vspace{-2mm}

\paragraph{Datasets}
We train OperatorNet on two types of datasets: humans and animals. 
For human shapes, our training set consists of 9440 shapes sampled from the DFAUST dataset~\cite{dfaust} and  8000 from the SURREAL dataset~\cite{varol17_surreal}, which is generated with the model proposed in~\cite{SMPL}.  The DFAUST dataset contains scan of human characters subject to a various of motions.  On the other hand, the SURREAL dataset injects more variability to the body types. 
 
For animals, we use the parametric model proposed in SMAL~\cite{smal} to generate 1800 animals of 3 different species -- lions, dogs, and horses. The meshes of the humans (resp. animals) are simplified to 1000 vertices (resp. 1769 vertices). 
\vspace{-4mm}

\paragraph{Input Shape Differences}
We construct the input shape differences using a truncated eigenbasis of dimension $60$ on the base shape, and the full basis on the target one, in all experiments, regardless of the number of vertices on the shapes.  The functional maps from the base to the targets are induced by the identity maps, since our training shapes are in 1-1 correspondence.  This implies that each of the shapes is represented by three $60\times 60$ matrices, representing the area-based, conformal and extrinsic shape differences respectively. 
The independence among the shape differences allows flexibility in selecting the combination of input shape differences, in Section~\ref{sec:result} we compare the performance of several combinations, and present a more detailed ablation study in Appendix~\ref{sec:abl}. 

It is worth noting that recent learning-based shape matching techniques enable efficient (functional) maps estimation. In particular, we use the unsupervised matching method of~\cite{USFMap} and evaluate OperatorNet trained with \emph{computed} shape differences in Section~\ref{sec:result}.

\vspace{-4mm}

\paragraph{Loss Function}
OperatorNet reconstructs coordinate functions of a given training shape. Our shape reconstruction loss operates in two steps. First, we estimate the optimal rigid transformation to align the ground truth point cloud $X_{\mbox{gt}}$ and the reconstructed point cloud $X_{\mbox{recon}}$ using the Kabsch algorithm~\cite{arun1987least} with ground truth correspondences. Secondly, we estimate the mean squared error between the aligned reconstruction and the ground truth.
\begin{equation}\label{equ:ICP_dist}
   L(X_{\mbox{gt}}, X_{\mbox{recon}}) =\frac {1}{n_V}\sum_{i=1}^{n_V}\Vert R( X_{\mbox{recon}}^{i})-{{X_{\mbox{gt}}^{i}}}\Vert^{2}.
\end{equation}
Here $R$ is the function that computes the optimal transformation between $X_{\mbox{recon}}$ and $X_{\mbox{gt}}$. We align the computed reconstruction to the ground truth embedding, so that the quality of the reconstructed point cloud is invariant to rigid transformations. This is important since the shape difference operators are invariant to rigid motion of the shape, and thus the network should not be penalized, for not recovering the correct orientation. 
On the other hand, this loss function is differentiable, since we use a closed-form expression of $R_{X_{\mbox{gt}}}$, given by the SVD, which enables back-propagation in neural network training.

\section{Evaluation}\label{sec:result}
In this section, we provide both qualitative and quantitative evaluations of the results from OperatorNet, and compare them to the geometric baselines. 

\vspace{-2mm}
\paragraph{Evaluation Metrics}
We denote by $S_{\mbox{gt}}$ and $S_{\mbox{recon}}$ the ground-truth and the reconstructed meshes respectively.
We let $d_R = L(X_{\mbox{gt}}, X_{\mbox{recon}})$, where $L$ is the rotationally-invariant distance defined in Eq.~\eqref{equ:ICP_dist} and $X$ is the vertex set of $S$.
Since OperatorNet is trained with the loss defined in Eq.~\eqref{equ:ICP_dist}, we introduce the following new metrics for a comprehensive, unbiased evaluation and comparison: (1) $d_V = |V(S_{\mbox{gt}}) - V(S_{\mbox{recon}})|/V(S_{\mbox{gt}})$, i.e., the relative error of mesh volumes; (3) $d_E = \mbox{mean}_{(i, j)} |l^{\mbox{gt}}_{ij} - l^{\mbox{recon}}_{ij}|/ l^{\mbox{gt}}_{ij}$, where $l_{ij}$ is the length of edge $(i, j)$. 

\vspace{-2mm}

\paragraph{Baselines} 
Two major baselines are considered: (1) the intrinsic reconstruction method from~\cite{boscaini2015shape}, in which we evaluate with the `Shape-from-Laplacian' option and use the full basis in \emph{both} the base shape and the target shape; (2) the reconstruction method from~\cite{Corman2017}, where the authors construct offset surfaces that also capture extrinsic geometry. Moreover, this method also provides a purely intrinsic reconstruction version. We evaluate both cases with the same basis truncation as our input. 
Beyond that, we also consider the nearest neighbor retrieval from the training set with respect to distances between shape difference matrices. 

\begin{figure}[t!]
  \centering
  \includegraphics[width=0.95\linewidth]{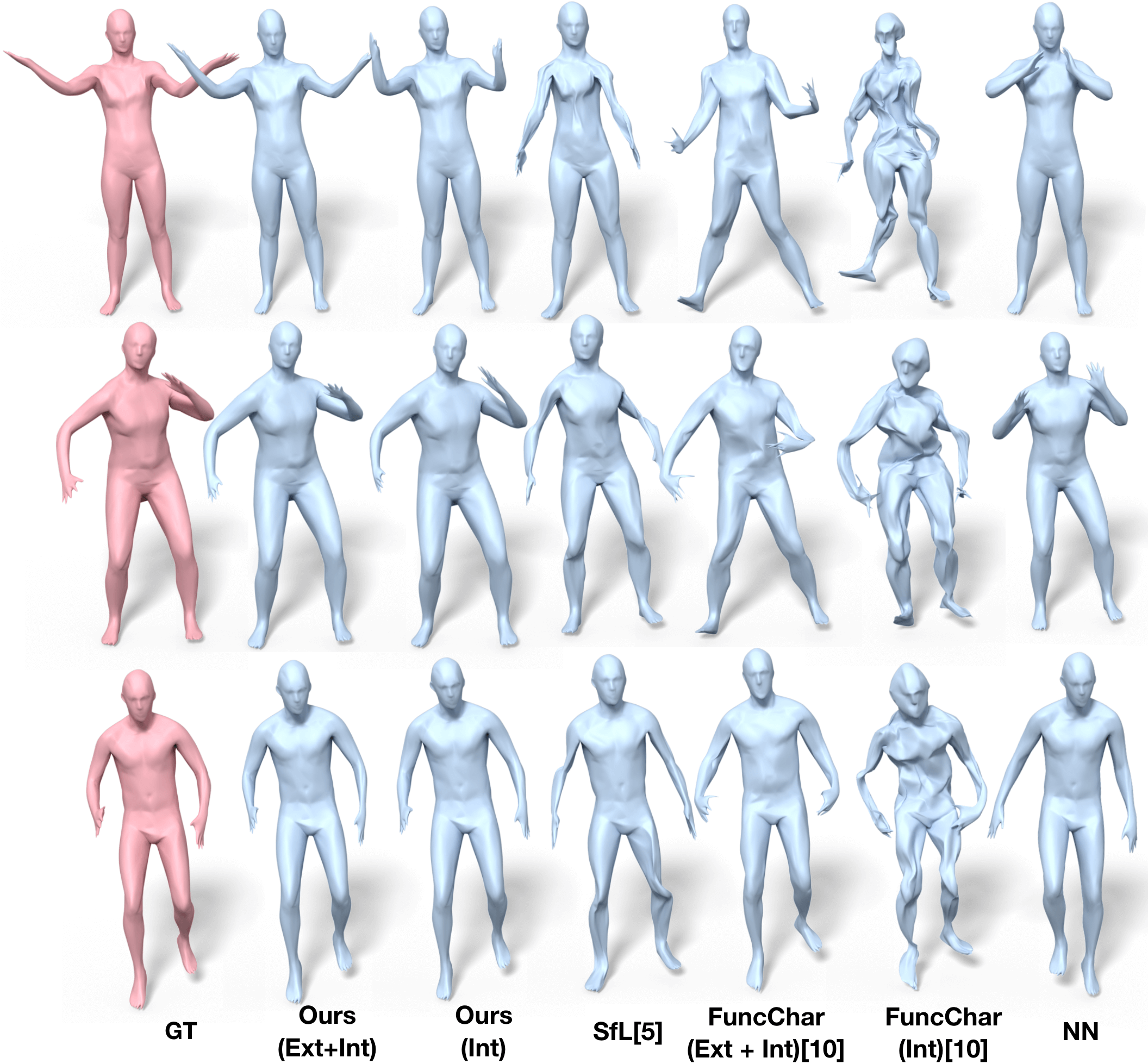}
\caption{\label{fig:qualitative} Qualitative comparison of our method and the baselines. }
\end{figure}

\vspace{-2mm}

\paragraph{Test Data}
We use 800 shapes from the DFAUST dataset as the test set, which contains 10 sub-collections (character + action sequence, each consisting of 80 shapes) that are isolated from the training/validation set. 
For the efficiency of baseline evaluation, we further sample 5 shapes via furthest point sampling regarding the pair-wise Hausdorff distance from each of the sub-collection, resulting in a set of 50 shapes that covers significant variability in both styles and poses in the test set.

 \vspace{-2.5mm}
 \paragraph{Qualitative Results}
 We demonstrate the reconstructed shapes from OperatorNet and the aforementioned baselines in Figure~\ref{fig:qualitative}, where the red shape in each row is the ground truth target shape. 
 The base shape in this experiment (also the base shape we compute shape differences on) is shown in Figure~\ref{fig:eigenfuncs}, which is in the rest pose. The geometric baselines in general perform worse under significant pose changes from the base (see the top two rows in Figure~\ref{fig:qualitative}), but give relatively more stable results when the difference is mainly in the shape style (see the bottom row). 
 
 Our method, on the other hand, produces consistently good reconstructions in all cases. Note also that, as expected, OperatorNet using all $3$ types of shape differences gives both the best quantitative and qualitative results. We provide more reconstruction examples in Appendix~\ref{sec:recon} highlighting the generalization power of our method.

 \vspace{-2.5mm} 
\paragraph{Quantitative Results}
We report all the quantitative metrics defined above in Table~\ref{table:quantitative_comparison}. 
First, we observe that OperatorNet using both intrinsic and extrinsic shape differences achieves the lowest reconstruction error, while the purely extrinsic version is the second best. 
Secondly, OperatorNet trained on shape differences from computed functional maps achieves competing performances, showing that our method is efficient even in the absence of ground truth bijective correspondences. 
Lastly, all the versions of OperatorNet significantly outperform the baselines.

Regarding the volume and edge recovery accuracy, either complete or intrinsic-only versions of OperatorNet achieve second to the best result. 
We remark that since the nearest neighbor search in general retrieves the right body type, therefore the volume is well-recovered. On the other hand, since the \emph{full} Laplacian is provided as input for the Shape-from-Laplacian baseline, it is expected to preserve intrinsic information. 

 \begin{table}[]
 \caption{Quantitative evaluation of shape reconstruction ($d_R$ is at the scale of $10^{-4}$).}
 \begin{tabular}{c|c|c|c}
 & $d_{R}$  & $d_V$ & $d_E$ \\ 
       \hline
       \hline
 \mbox{Op.Net (Int+Ext) }  & $\mathbf{1.11}$ & 0.014 & 0.045 \\
 \mbox{Op.Net (Int)}   & 2.41      &   0.013    & 0.046 \\
 \mbox{Op.Net (Ext)}   & 1.25      &    0.017   & 0.046 \\
 \mbox{Op.Net (Comp)(Ext)}  & 3.86 &  0.021 & 0.052 \\
 \mbox{Op.Net (Comp)(Int+Ext)} & 6.22  & 0.022  & 0.053 \\
 
 \mbox{SfL }\cite{boscaini2015shape}      & 48.8 & 0.081 & $\mathbf{0.012}$ \\
 \mbox{FuncChar }\cite{Corman2017}\mbox{(Int) }      & 65.1 & 0.356 & 0.118 \\
 \mbox{FuncChar }\cite{Corman2017} \mbox{(Int+Ext)} & 28.4 & 0.028 & 0.110 \\
 \mbox{NN}    &   25.5   &    $\mathbf{0.005}$   &  0.043     \\
 \end{tabular}
 \label{table:quantitative_comparison}
 \vspace{-2mm}
 \end{table}

\begin{figure}[t!]
    \centering
     \includegraphics[width=\columnwidth]{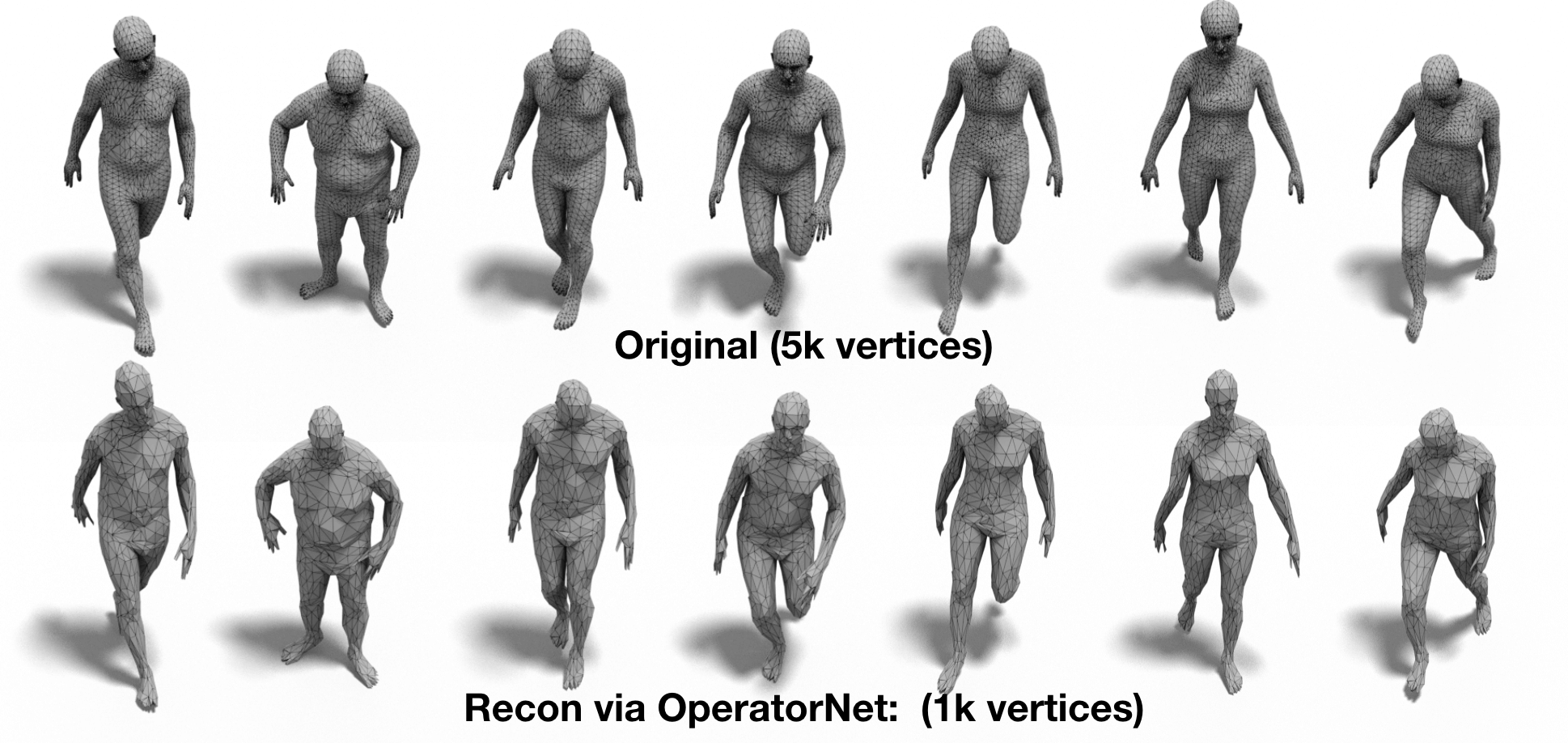}
    \caption{Top row: input shapes with different number of vertices than that of the base shape; Bottom row: reconstructions via OperatorNet. 
    \vspace{-2mm}
    \label{fig:recon_diff_tri}}
\end{figure}

\vspace{-3mm}
\paragraph{Reconstructions of Shapes with Different Discretizations}
Lastly, we show that our approach is capable of encoding differences between shapes with different discretizations. 
In Figure~\ref{fig:recon_diff_tri}, we compute the functional maps from the fine meshes (top row, with 5k vertices) by projecting them to a lower resolution base mesh with 1k vertices. We then reconstruct them with OperatorNet trained on lower resolution shapes. 
This, on the other hand, is extremely difficult for purely geometric methods. In Appendix~\ref{sec:recon} we provide examples of reconstructions in the same setting using the method of~\cite{Corman2017}, and reconstructions with OperatorNet trained with shapes having 2k vertices.

\section{Applications}\label{sec:app}
In this section, we present all of our results using OperatorNet trained with all $3$ types of shape differences.

\vspace{-2mm}
\paragraph{Shape Interpolation}
Given two shapes, we first interpolate their shape differences using the formulation in Eq.\eqref{eq:interp}, and then synthesize intermediate shapes by inferring the interpolated shape differences 
with OperatorNet. 

\begin{figure*}[h!]
   \centering
   \includegraphics[width=0.9\linewidth]{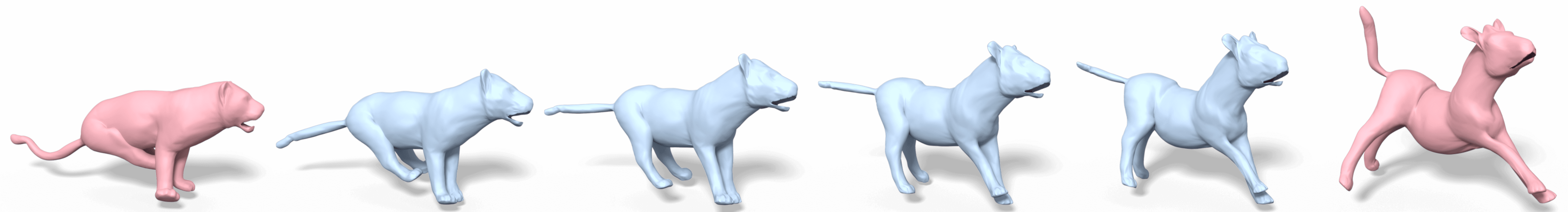}
 \caption{\label{fig:interpolation_animals} Shape interpolation from a tiger (left) to a horse (right) using OperatorNet trained on animals dataset.}
 \vspace{-4mm}
\end{figure*}

\begin{figure}[t!]
   \centering
   \includegraphics[width=0.9\linewidth]{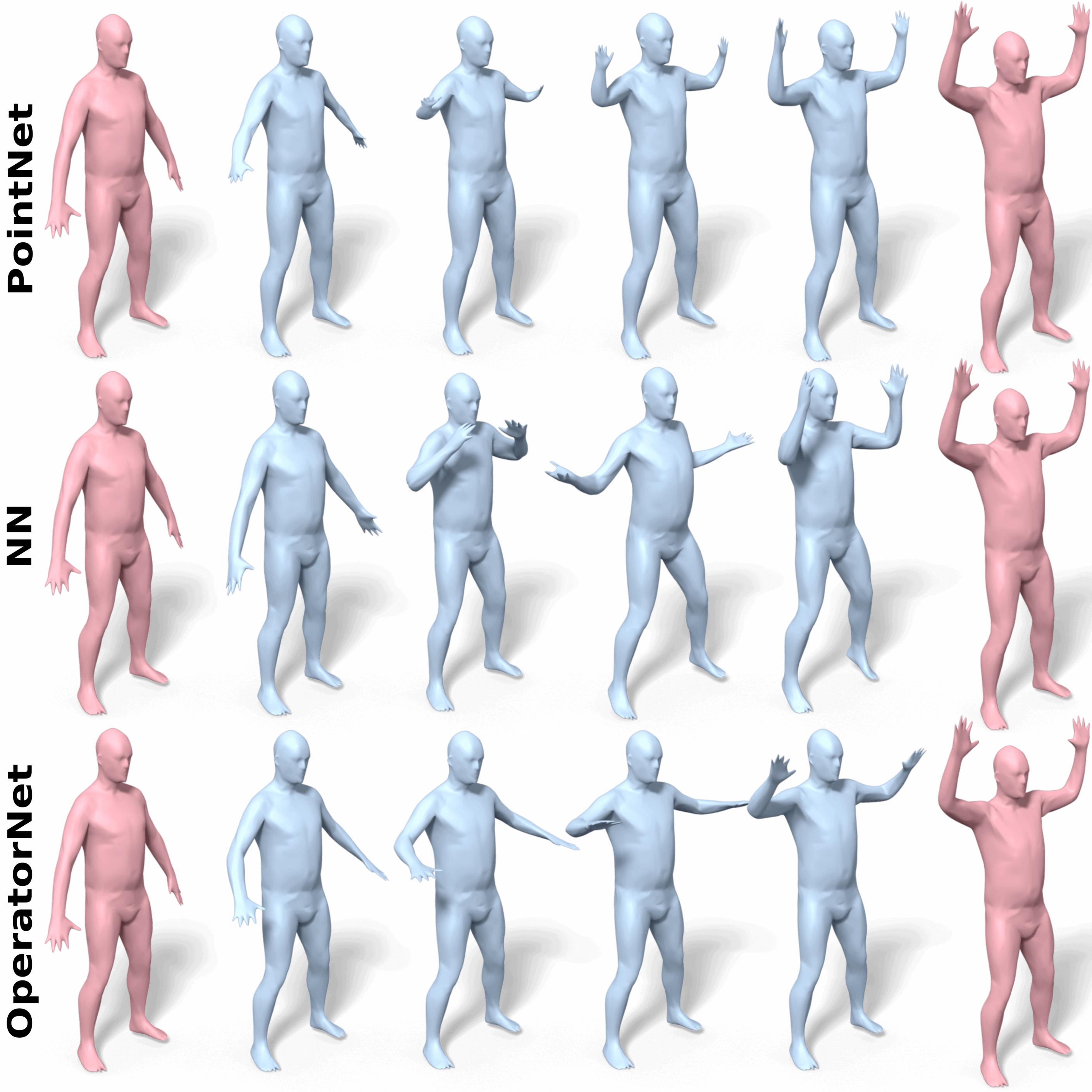}
 \caption{\label{fig:interpolation}Shape interpolation between two humans. Note that PointNet autoencoder produces shapes with local area distortion, while the interpolation from nearest neighbor (NN) retrieval is not continuous. }
\end{figure}

We compare our method against nearest neighbor retrieval and PointNet autoencoder. 
PointNet autoencoder is trained with the encoder architecture from \cite{qi2016_pointnet} and with our decoder. Two versions of PointNet are trained: one autoencoder with spatial transformers and one without. Since the autoencoder without spatial transformers performs better in our experiments, we select it for the comparisons. 
Nearest neighbor interpolation retrieves the nearest neighbor of the interpolated shape differences in the training set and uses the corresponding embedding.  
As expected, (see the second row of Figure~\ref{fig:interpolation}), nearest neighbor interpolation is less continuous. 

As shown in Figure~\ref{fig:teaser_interpolation}, our method produces smooth interpolations, without significant local area distortions compared to PointNet.
Similarly, in Figure~\ref{fig:interpolation}, we observe that the interpolation via PointNet suffers from local distortion on the arms. In contrast, interpolation using OperatorNet is continuous and respects the structure and constraints of the body, suggesting that shape differences efficiently encode the shape structure. 
We provide further comparisons to other baselines including~\cite{PNPP, ben2018multi, groueix2018b} and to linear interpolation of shape differences in Appendix~\ref{sec:interp}.

We also train OperatorNet on the animals dataset as described in Section~\ref{sec:network} and show in Figure~\ref{fig:interpolation_animals}  an interpolation from a tiger to a horse.

\vspace{-4mm}

\paragraph{Shape Analogy}
Our second application is to construct semantically meaningful new shapes based on shape analogies. 
Given shapes $S_A, S_B, S_C$, our goal is to construct a new shape $S_X$, such that $S_C$ relates to $S_X$ as $S_A$ to $S_B$. 

Following the discussion in Section~\ref{sec:background}, the functoriality of shape differences allows an explicit and mathematically meaningful way of constructing the shape difference of $S_X$, given that of $S_A, S_B$ and $S_C$. Namely, $\D_{X} = \D_{C} \D_{A}^{+} \D_{B}. $
Then, with our OperatorNet, we reconstruct the embedding of the unknown $S_X$ by feeding $\D_{X}$ to the network. 

We compare our results to that of the PointNet autoencoder. 
In the latter, we reconstruct $S_X$ by decoding the latent code obtained by $l_X = l_C - l_A + l_B$, where $l_A$ is the latent code of shape $S_A$ (and similarly for $S_B, S_C$).  

\begin{figure}[t]
   \centering
   \includegraphics[width=0.95\linewidth]{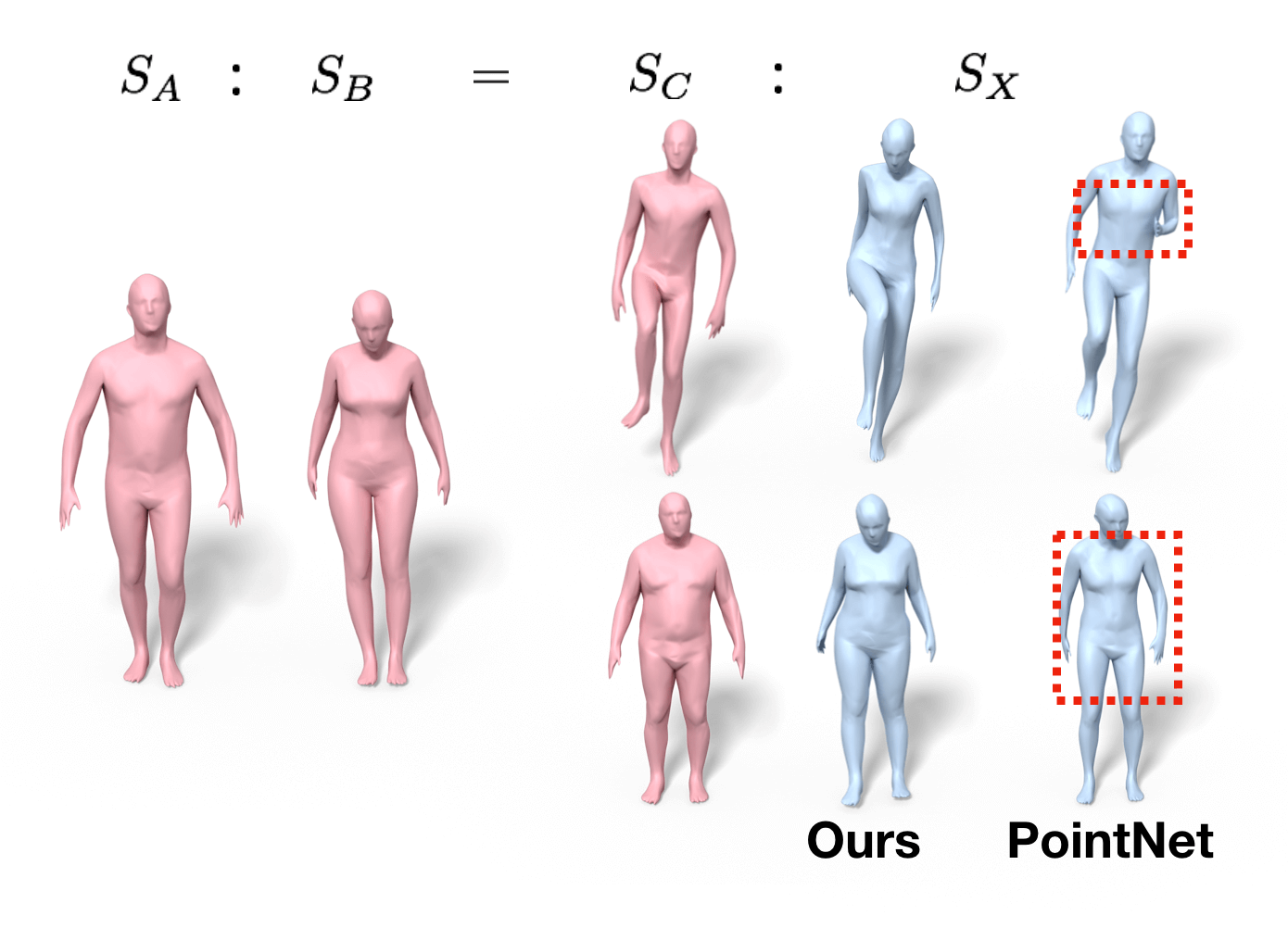}
 \caption{\label{fig:analogy}  Transferring gender via shape analogies: $S_A$ and $S_B$ are a fixed pair of human shapes with similar poses and styles, but of different genders. We generate $S_X$, which is supposed to be a ``female'' version of the varying $S_C$. Our analogies are semantically meaningful, while PointNet can produce suboptimal results (see the red dotted boxes for the discrepancies).}
 \vspace{-2mm}
\end{figure}

In Figure~\ref{fig:analogy}, we show a set of shape analogies obtained via OperatorNet and PointNet autoencoder. It is evident that our results are both more natural and intuitive. We also refer the readers to Appendix~\ref{sec:analogy} for more examples of analogies. 

\section{Conclusion \& Future Work}
In this paper we have introduced a novel learning-based technique for recovering shapes from their difference operators. Our key observation is that shape differences, stored as compact matrices lend themselves naturally to learning and allow to both recover the underlying shape space in a collection and encode the geometry of individual shapes. We also introduce a novel extrinsic shape difference operator and show its utility for shape reconstruction and other applications such as shape interpolation and analogies.

Currently our approach is only well-adapted to shapes represented as triangle meshes. Thus, in the future we plan to extend this framework to both learn the optimal inner products from data, and adapt our pipeline to other shape representations, such as point clouds or triangle soups.

\vspace{-3mm}
\paragraph{Acknowledgements}
	Parts of this work were supported by the ERC Starting Grant
        StG-2017-758800 (EXPROTEA), KAUST OSR Award CRG-2017-3426 a
        gift from the Nvidia Corporation, a
        Vannevar Bush Faculty Fellowship, NSF grant DMS-1546206, a
        Google Research award and gifts from Adobe and Autodesk.
        The authors thank Davide Boscaini and Etienne Corman for their help with baseline comparisons. 
        {\small
\bibliographystyle{ieee_fullname}
\bibliography{egbib}

\begin{thebibliography}{10}\itemsep=-1pt

\bibitem{achlioptas2017learning}
Panos Achlioptas, Olga Diamanti, Ioannis Mitliagkas, and Leonidas Guibas.
\newblock Learning representations and generative models for 3d point clouds.
\newblock {\em arXiv preprint arXiv:1707.02392}, 2017.

\bibitem{scape}
Dragomir Anguelov, Praveen Srinivasan, Daphne Koller, Sebastian Thrun, Jim
  Rodgers, and James Davis.
\newblock {SCAPE}: {S}hape {C}ompletion and {A}nimation of {P}eople.
\newblock In {\em ACM Transactions on Graphics (TOG)}, volume~24, pages
  408--416. ACM, 2005.

\bibitem{ben2018multi}
Heli Ben-Hamu, Haggai Maron, Itay Kezurer, Gal Avineri, and Yaron Lipman.
\newblock Multi-chart generative surface modeling.
\newblock In {\em Proc. SIGGRAPH Asia}, page 215. ACM, 2018.

\bibitem{dfaust}
Federica Bogo, Javier Romero, Gerard Pons-Moll, and Michael~J. Black.
\newblock Dynamic {FAUST}: {R}egistering human bodies in motion.
\newblock In {\em CVPR}, July 2017.

\bibitem{boscaini2015shape}
Davide Boscaini, Davide Eynard, Drosos Kourounis, and Michael~M Bronstein.
\newblock Shape-from-operator: Recovering shapes from intrinsic operators.
\newblock In {\em Computer Graphics Forum}, volume~34, pages 265--274. Wiley
  Online Library, 2015.

\bibitem{boscaini2016learning}
Davide Boscaini, Jonathan Masci, Emanuele Rodol{\`a}, and Michael Bronstein.
\newblock Learning shape correspondence with anisotropic convolutional neural
  networks.
\newblock In {\em Advances in Neural Information Processing Systems}, pages
  3189--3197, 2016.

\bibitem{bronstein2017geometric}
Michael~M Bronstein, Joan Bruna, Yann LeCun, Arthur Szlam, and Pierre
  Vandergheynst.
\newblock Geometric deep learning: going beyond euclidean data.
\newblock {\em IEEE Signal Processing Magazine}, 34(4):18--42, 2017.

\bibitem{shapenet2015}
Angel~X. Chang, Thomas~A. Funkhouser, Leonidas~J. Guibas, Pat Hanrahan,
  Qi{-}Xing Huang, Zimo Li, Silvio Savarese, Manolis Savva, Shuran Song, Hao
  Su, Jianxiong Xiao, Li Yi, and Fisher Yu.
\newblock Shapenet: An information-rich 3d model repository.
\newblock {\em CoRR}, abs/1512.03012, 2015.

\bibitem{corman2016}
Etienne Corman.
\newblock {\em Functional representation of deformable surfaces for geometry
  processing}.
\newblock PhD thesis, 2016.
\newblock PhD thesis.

\bibitem{Corman2017}
Etienne Corman, Justin Solomon, Mirela Ben-Chen, Leonidas Guibas, and Maks
  Ovsjanikov.
\newblock Functional characterization of intrinsic and extrinsic geometry.
\newblock {\em ACM Trans. Graph.}, 36(2):14:1--14:17, Mar. 2017.

\bibitem{MDS}
Trevor~F. Cox and M.A.A. Cox.
\newblock {\em Multidimensional Scaling, Second Edition}.
\newblock Chapman and Hall/CRC, 2 edition, 2000.

\bibitem{groueix2018b}
Thibault Groueix, Matthew Fisher, Vladimir~G. Kim, Bryan Russell, and Mathieu
  Aubry.
\newblock 3d-coded : 3d correspondences by deep deformation.
\newblock In {\em ECCV}, 2018.

\bibitem{hasler2009statistical}
Nils Hasler, Carsten Stoll, Martin Sunkel, Bodo Rosenhahn, and H-P Seidel.
\newblock A statistical model of human pose and body shape.
\newblock In {\em Computer graphics forum}, volume~28, pages 337--346. Wiley
  Online Library, 2009.

\bibitem{limitshape}
Ruqi Huang, Panos Achlioptas, Leonidas Guibas, and Maks Ovsjanikov.
\newblock {Limit Shapes - A Tool for Understanding Shape Differences and
  Variability in 3D Model Collections}.
\newblock {\em Computer Graphics Forum}, 2019.

\bibitem{kovnatsky2013coupled}
Artiom Kovnatsky, Michael~M Bronstein, Alexander~M Bronstein, Klaus Glashoff,
  and Ron Kimmel.
\newblock Coupled quasi-harmonic bases.
\newblock In {\em Computer Graphics Forum}, volume~32, pages 439--448. Wiley
  Online Library, 2013.

\bibitem{levy2006}
Bruno Levy.
\newblock Laplace-beltrami eigenfunctions towards an algorithm that
  "understands" geometry.
\newblock In {\em IEEE International Conference on Shape Modeling and
  Applications 2006 (SMI'06)}, pages 13--13, June 2006.

\bibitem{litany2018deformable}
Or Litany, Alex Bronstein, Michael Bronstein, and Ameesh Makadia.
\newblock Deformable shape completion with graph convolutional autoencoders.
\newblock In {\em Proc. CVPR}, pages 1886--1895, 2018.

\bibitem{litany2017deep}
Or Litany, Tal Remez, Emanuele Rodol{\`a}, Alex Bronstein, and Michael
  Bronstein.
\newblock Deep functional maps: Structured prediction for dense shape
  correspondence.
\newblock In {\em Proc. ICCV}, pages 5659--5667, 2017.

\bibitem{SMPL}
Matthew Loper, Naureen Mahmood, Javier Romero, Gerard Pons-Moll, and Michael~J.
  Black.
\newblock Smpl: A skinned multi-person linear model.
\newblock {\em ACM Trans. Graph.}, 34(6):248:1--248:16, Oct. 2015.

\bibitem{maron2017convolutional}
Haggai Maron, Meirav Galun, Noam Aigerman, Miri Trope, Nadav Dym, Ersin Yumer,
  Vladimir Kim, and Yaron Lipman.
\newblock Convolutional neural networks on surfaces via seamless toric covers.
\newblock 2017.

\bibitem{masci2015geodesic}
Jonathan Masci, Davide Boscaini, Michael Bronstein, and Pierre Vandergheynst.
\newblock Geodesic convolutional neural networks on riemannian manifolds.
\newblock In {\em Proc. ICCV workshops}, pages 37--45, 2015.

\bibitem{maturana2015voxnet}
Daniel Maturana and Sebastian Scherer.
\newblock Voxnet: A 3d convolutional neural network for real-time object
  recognition.
\newblock In {\em Intelligent Robots and Systems (IROS), 2015 IEEE/RSJ
  International Conference on}, pages 922--928. IEEE, 2015.

\bibitem{meyer03}
Mark Meyer, Mathieu Desbrun, Peter Schr{\"o}der, and Alan~H Barr.
\newblock {D}iscrete {D}ifferential-{G}eometry {O}perators for {T}riangulated
  2-{M}anifolds.
\newblock In {\em Visualization and mathematics III}, pages 35--57. Springer,
  2003.

\bibitem{Functional}
Maks Ovsjanikov, Mirela Ben-Chen, Justin Solomon, Adrian Butscher, and Leonidas
  Guibas.
\newblock {F}unctional {M}aps: {A} {F}lexible {R}epresentation of {M}aps
  {B}etween {S}hapes.
\newblock {\em ACM Transactions on Graphics (TOG)}, 31(4):30, 2012.

\bibitem{ovsjanikov2017computing}
Maks Ovsjanikov, Etienne Corman, Michael Bronstein, Emanuele Rodol{\`a}, Mirela
  Ben-Chen, Leonidas Guibas, Frederic Chazal, and Alex Bronstein.
\newblock Computing and processing correspondences with functional maps.
\newblock In {\em ACM SIGGRAPH 2017 Courses}, 2017.

\bibitem{pinkall1993computing}
Ulrich Pinkall and Konrad Polthier.
\newblock {C}omputing {D}iscrete {M}inimal {S}urfaces and their {C}onjugates.
\newblock {\em Experimental mathematics}, 2(1):15--36, 1993.

\bibitem{poulenard2018multi}
Adrien Poulenard and Maks Ovsjanikov.
\newblock Multi-directional geodesic neural networks via equivariant
  convolution.
\newblock {\em ACM Trans. Graph. (Proc. SIGGRAPH Asia)}, 37(6):236:1--236:14,
  2018.

\bibitem{qi2016_pointnet}
Charles~Ruizhongtai Qi, Hao Su, Kaichun Mo, and Leonidas~J. Guibas.
\newblock Pointnet: deep learning on point sets for 3d classification and
  segmentation.
\newblock {\em CoRR}, abs/1612.00593, 2016.

\bibitem{qi2016volumetric}
Charles~R Qi, Hao Su, Matthias Nie{\ss}ner, Angela Dai, Mengyuan Yan, and
  Leonidas~J Guibas.
\newblock Volumetric and multi-view cnns for object classification on 3d data.
\newblock In {\em Proc. CVPR}, pages 5648--5656, 2016.

\bibitem{PNPP}
Charles~Ruizhongtai Qi, Li Yi, Hao Su, and Leonidas~J. Guibas.
\newblock Pointnet++: Deep hierarchical feature learning on point sets in a
  metric space.
\newblock {\em CoRR}, abs/1706.02413, 2017.

\bibitem{USFMap}
Jean{-}Michel Roufosse, Abhishek Sharma, and Maks Ovsjanikov.
\newblock Unsupervised deep learning for structured shape matching.
\newblock {\em CoRR}, abs/1812.03794, 2018.

\bibitem{Rustamov2013}
Raif~M. Rustamov, Maks Ovsjanikov, Omri Azencot, Mirela Ben-Chen,
  Fr{\'{e}}d{\'{e}}ric Chazal, and Leonidas Guibas.
\newblock {Map-based exploration of intrinsic shape differences and
  variability}.
\newblock {\em ACM Transactions on Graphics (TOG)}, 32(4):1, 2013.

\bibitem{Schulz2017}
Adriana Schulz, Ariel Shamir, Ilya Baran, David I.~W. Levin, Pitchaya
  Sitthi-Amorn, and Wojciech Matusik.
\newblock Retrieval on parametric shape collections.
\newblock {\em ACM Trans. Graph.}, 36(4), Jan. 2017.

\bibitem{sinha2016deep}
Ayan Sinha, Jing Bai, and Karthik Ramani.
\newblock Deep learning 3d shape surfaces using geometry images.
\newblock In {\em European Conference on Computer Vision}, pages 223--240.
  Springer, 2016.

\bibitem{Surfnet}
Ayan Sinha, Asim Unmesh, Qixing Huang, and Karthik Ramani.
\newblock Surfnet: Generating 3d shape surfaces using deep residual networks.
\newblock In {\em The IEEE Conference on Computer Vision and Pattern
  Recognition (CVPR)}, July 2017.

\bibitem{arun1987least}
Arun~K. Somani, Thomas~S. Huang, and Steven~D. Blostein.
\newblock Least-squares fitting of two 3-d point sets.
\newblock {\em IEEE Transactions on pattern analysis and machine intelligence},
  (5):698--700, 1987.

\bibitem{varol17_surreal}
G{\"u}l Varol, Javier Romero, Xavier Martin, Naureen Mahmood, Michael~J. Black,
  Ivan Laptev, and Cordelia Schmid.
\newblock Learning from synthetic humans.
\newblock In {\em CVPR}, 2017.

\bibitem{wang2017cnn}
Peng-Shuai Wang, Yang Liu, Yu-Xiao Guo, Chun-Yu Sun, and Xin Tong.
\newblock O-cnn: Octree-based convolutional neural networks for 3d shape
  analysis.
\newblock {\em ACM Transactions on Graphics (TOG)}, 36(4):72, 2017.

\bibitem{smal}
Silvia Zuffi, Angjoo Kanazawa, David Jacobs, and Michael~J. Black.
\newblock {3D} menagerie: Modeling the {3D} shape and pose of animals.
\newblock In {\em CVPR}, July 2017.

\end{thebibliography}
}

\appendix

\section{Proof of Theorem 1.}\label{sec:proof}
\begin{proof}
    Since $X$ is known to be of rank 3, and $\mathbf{G}$ is symmetric, we have, by SVD: 
    \[\mathbf{G} = \bPhi^T A X X^T A \bPhi = U \Sigma U^T, \]
    where, $U, \Sigma$ are respectively the top 3 singular vectors and singular values of $\mathbf{G}$. Therefore, we have $\bPhi^T A X R = U\sqrt{\Sigma}$, where $R$ is a $3\times 3$ rigid transformation matrix satisfying $R^T R = I_{3\times 3}$. 
    In other words, we recover $\bPhi^T A \tilde{X}$ from $E^G$, where $\tilde{X} = XR$ is equivalent to $X$ up to rigid transformations.
    Then, to recover the projection of $\tilde{X}$ in the space spanned by $\bPhi$, we simply compute $\bPhi \bPhi^T A \tilde{X}$.
\end{proof}

\section{Ablation Study on Network Design}\label{sec:abl}

We investigate multiple architectures for OperatorNet. In Table~\ref{table:ablation} we compare the reconstruction performance over different combinations of input shape differences, and different depths of encoders.

We report the performance of 4 different convolutional encoders from 1 to 4 layers deep by doubling the number of neurons every layer.

Two trends are observed in Table~\ref{table:ablation}: first, we always achieve the best performance when all three types of shape differences are used, for varying depths of the network; second, fixing the combination of input shape differences, the network performs better as its depth gets shallower. 

Putting these two observations together, we justify our final model, which has one single layer convolutional encoder and uses all three types of shape differences as input.

\begin{table*}[!htbp]
\centering
\caption{Ablation study: auto-encoder performance on DFAUST testset (measured by the loss function as defined in Eq~\eqref{equ:ICP_dist}, the errors in the table are at the scale of $10^{-4}$). }
\begin{tabular}{l|c|c|c|c|c|c|c}
Encoder architecture & Area & Ext & Conf & A+E & A+C & E+C & A+E+C \\
\hline
Conv. 8 & \textbf{8.61} & \textbf{4.29} & \textbf{3.78} & \textbf{3.82}& \textbf{3.41} & \textbf{2.56} & \textcolor{red}{\textbf{2.46}}\\
Conv. 8$\times$16 & 9.08 & 4.54 & 4.28 & 4.65 & 3.93 & 3.10 & \textbf{3.05}\\
Conv. 8$\times$16$\times$32 & 9.90 & 5.54 & 4.91 & 5.59 & 4.88 & 3.71 & \textbf{3.55}\\
Conv. 8$\times$16$\times$32$\times$64 & 11.16 & 6.39 & 5.93 & 6.89 & 5.42 & 4.35 & \textbf{4.24}\\
\end{tabular}
\label{table:ablation}
\end{table*}

\section{Shape Reconstructions}\label{sec:recon}
\begin{figure*}[!ht]
\centering
\begin{subfigure}{0.5\textwidth}
  \centering
  \includegraphics[width=\linewidth]{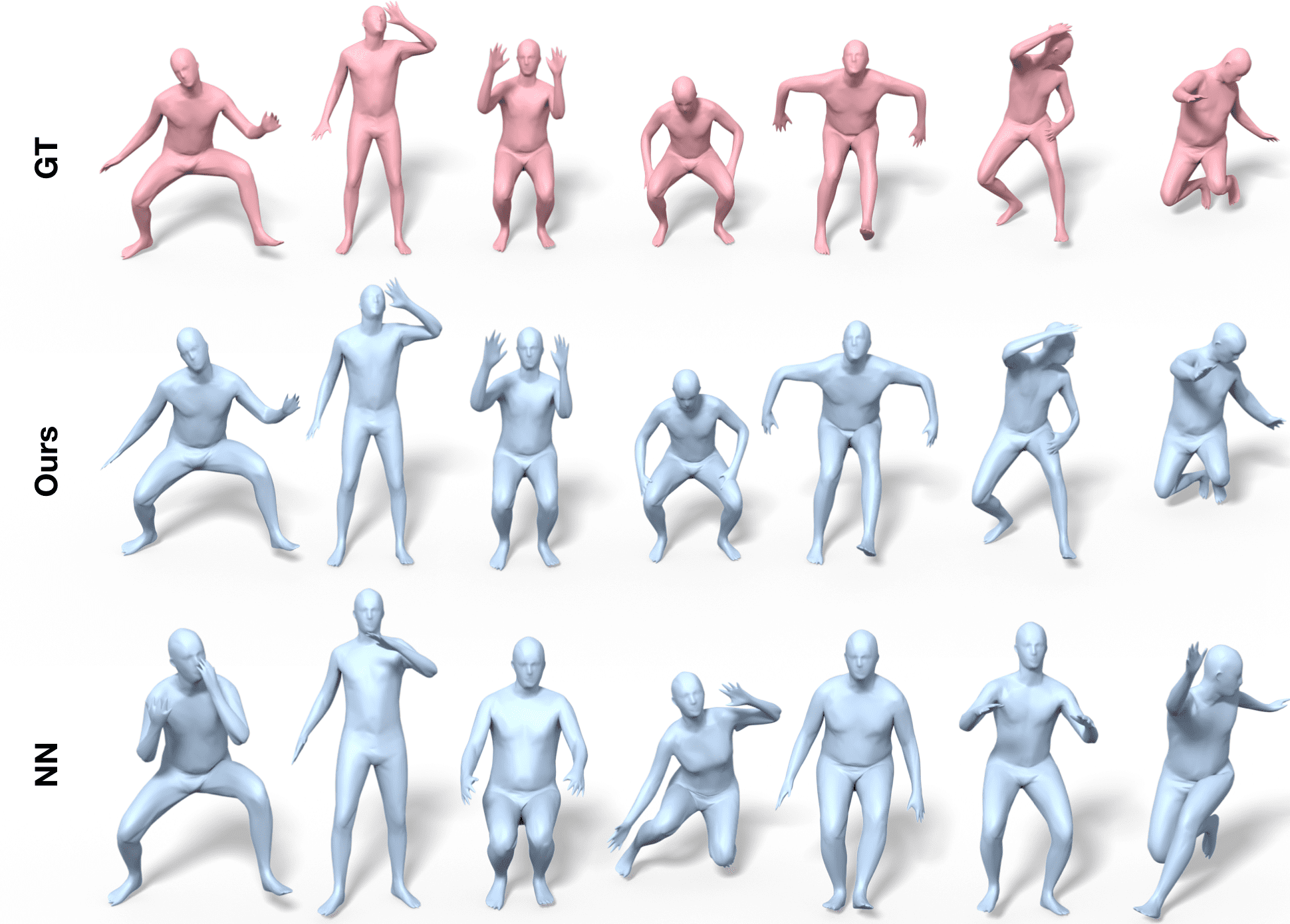}
\end{subfigure}%
\begin{subfigure}{0.5\textwidth}
  \centering
  \includegraphics[width=0.95\linewidth]{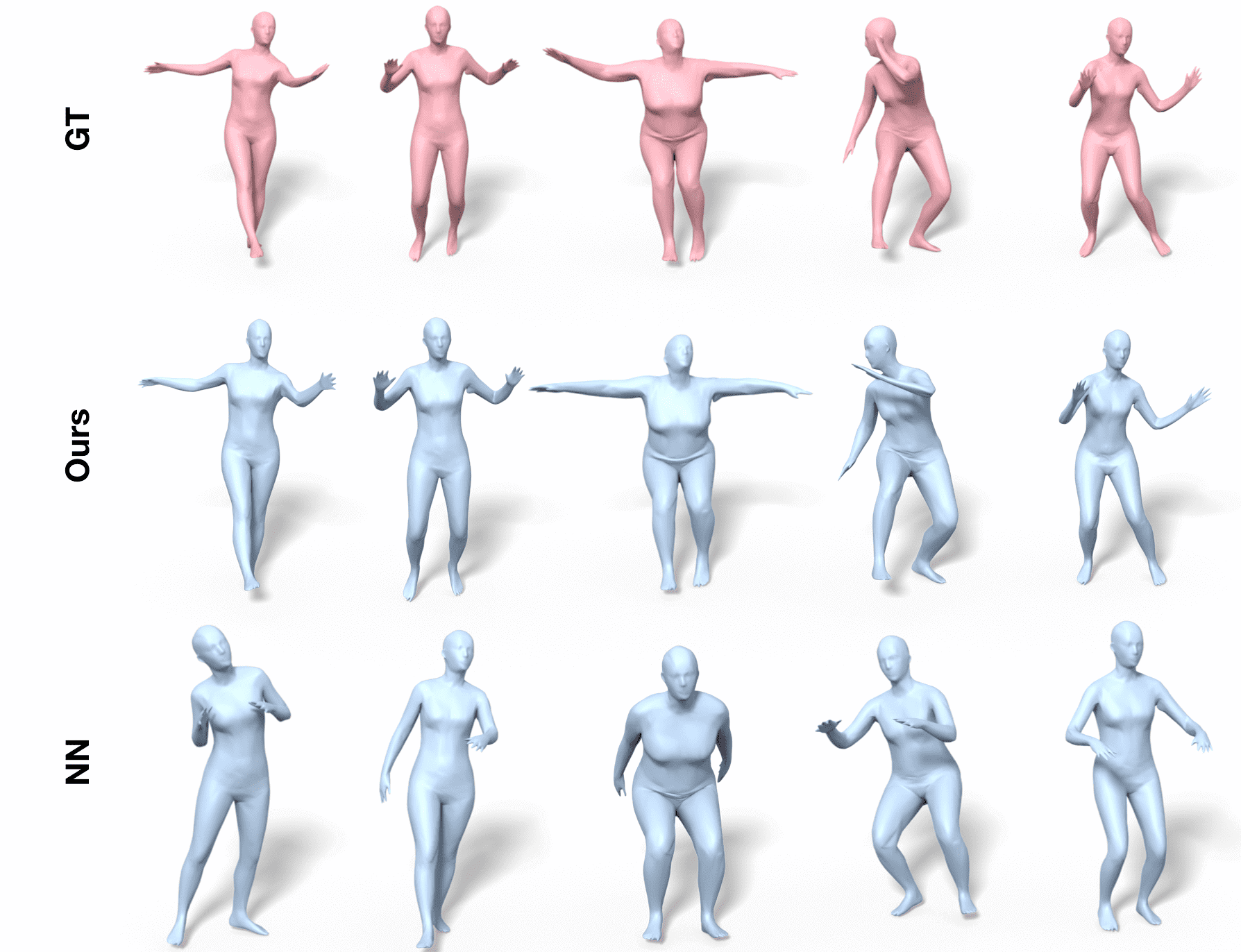}
\end{subfigure}
\caption{\label{fig:recon_male_female} Top row: ground-truth embeddings; middle row: reconstructions via OperatorNet; bottom row: shapes from the training set, whose shape differences that are closest to the ones of the test shapes in the top row.}
\end{figure*}

\paragraph{Verification of Generalization Power of OperatorNet} To demonstrate the generalization power of OperatorNet, we show in Figure~\ref{fig:recon_male_female} our reconstructions of test shapes from the SURREAL dataset.
For comparison, we retrieve the shapes in the training set, whose shape differences are the nearest to the ones of the test shapes. 
In each of the figures, the top row presents the ground-truth test shapes; the middle row shows reconstructions from OperatorNet; the bottom row demonstrates the shapes retrieved from the training set via nearest neighbor search in the space of shape differences.

It is evident that OperatorNet accurately reconstructs the test shapes, which deviate from the shapes in the training set significantly, suggesting that our network generalizes well in unseen data.

\begin{figure}[!t]
   \centering
   \includegraphics[width=\linewidth]{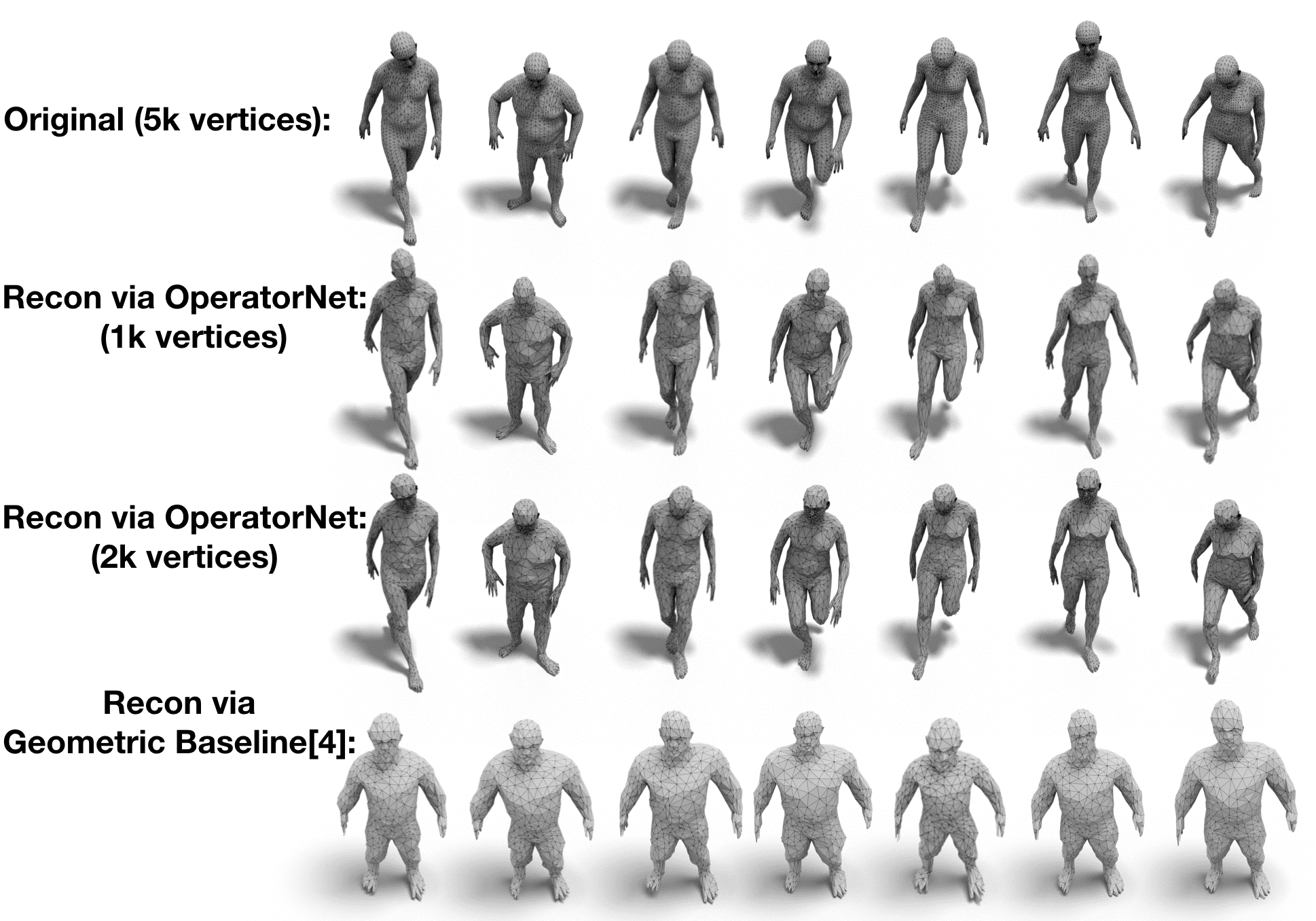}
 \caption{\label{fig:diff_tri} Top row: input shapes with different number of vertices  (5k) than that of the base shape (1k); second row: reconstructions of the original OperatorNet; third row: reconstructions of OperatorNet trained with higher resolution labels (2k vertices); bottom row: reconstructions via the geometric approach~\cite{Corman2017}.}
 \end{figure}

\paragraph{Reconstruction of Shapes in Different Discretizations} We show the reconstructions of shapes in a different discretization than the base shape in Figure~\ref{fig:diff_tri}, part of which (the top two rows) is demonstrated in Figure.~\ref{fig:recon_diff_tri}. Here we further train an OperatorNet with finer labels (of 2k vertices compared to that of 1k vertices used in the original version) and show the reconstructions on the third row of Figure~\ref{fig:diff_tri}. 
We emphasize that the use of coarse labels is for a fair comparison to the geometrical baselines for reconstructing embeddings from shape differences. As shown in the third row, OperatorNet reconstructs the shapes in a higher resolution well, which is not possible for the geometric approaches. 

Reconstructing shapes of different triangulations is extremely difficult for geometric approach: we demonstrate the reconstructions via the geometric approach~\cite{Corman2017} in the bottom row: the outputs are all close to the source shape (i.e., the base shape), which suggests that the algorithm struggles to find the right direction to deform the source to the target.

\section{Shape Analogies}\label{sec:analogy}
\begin{figure*}[!ht]
  \centering
  \includegraphics[width=0.8\linewidth]{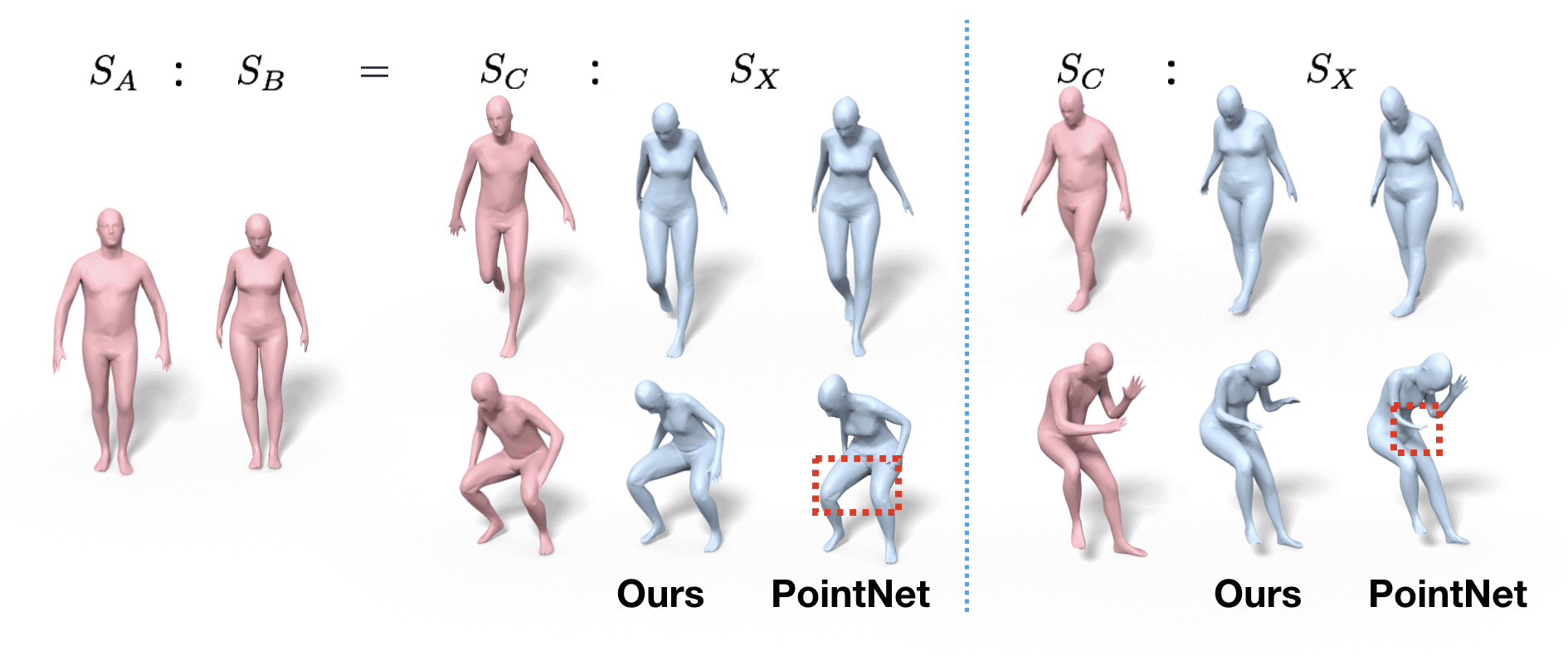}
\caption{\label{fig:gender_analogy} Gender analogies via OperatorNet and PointNet. Note that though in some cases PointNet also delivers reasonable results (e.g. the ones on the top row), the results of OperatorNet are more natural and semantically meaningful (see, e.g., the discrepancies highlighted in the red dotted boxes). } 
\end{figure*}

In addition to Figure~\ref{fig:analogy}, we present more gender analogies in Figure~\ref{fig:gender_analogy}. Note that though in some cases PointNet also delivers reasonable results (e.g. the ones on the top row), the results of OperatorNet are in general more natural and semantically meaningful (see, e.g., the discrepancies highlighted in the red dotted boxes).

\begin{figure}[!ht]
  \centering
  \includegraphics[width=\linewidth]{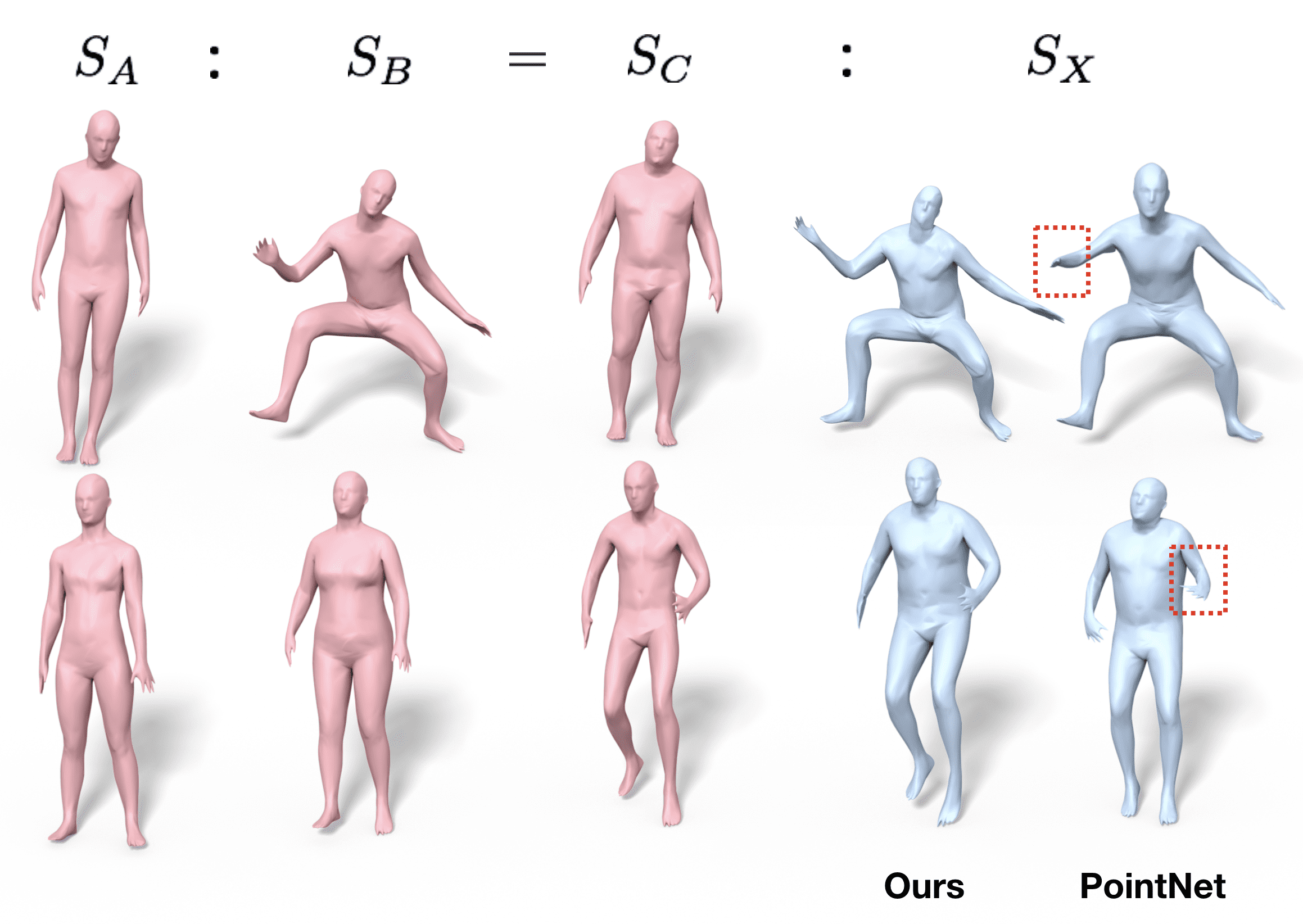}
\caption{\label{fig:shape_analogy}  Human shape analogies via OperatorNet and PointNet auto-encoder (see the red dotted boxes for the discrepancies).} 
\end{figure}

We also present a set of shape analogies that transfer pose (top row) and style (bottom row) across human shapes in Figure~\ref{fig:shape_analogy}. We observe that our results (the fourth column from the left) are both more natural and intuitive while PointNet (the right-most column) produces less satisfactory results with, e.g., local area distortions (see the red dotted boxes).

\begin{figure}[!ht]
  \centering
  \includegraphics[width=\linewidth]{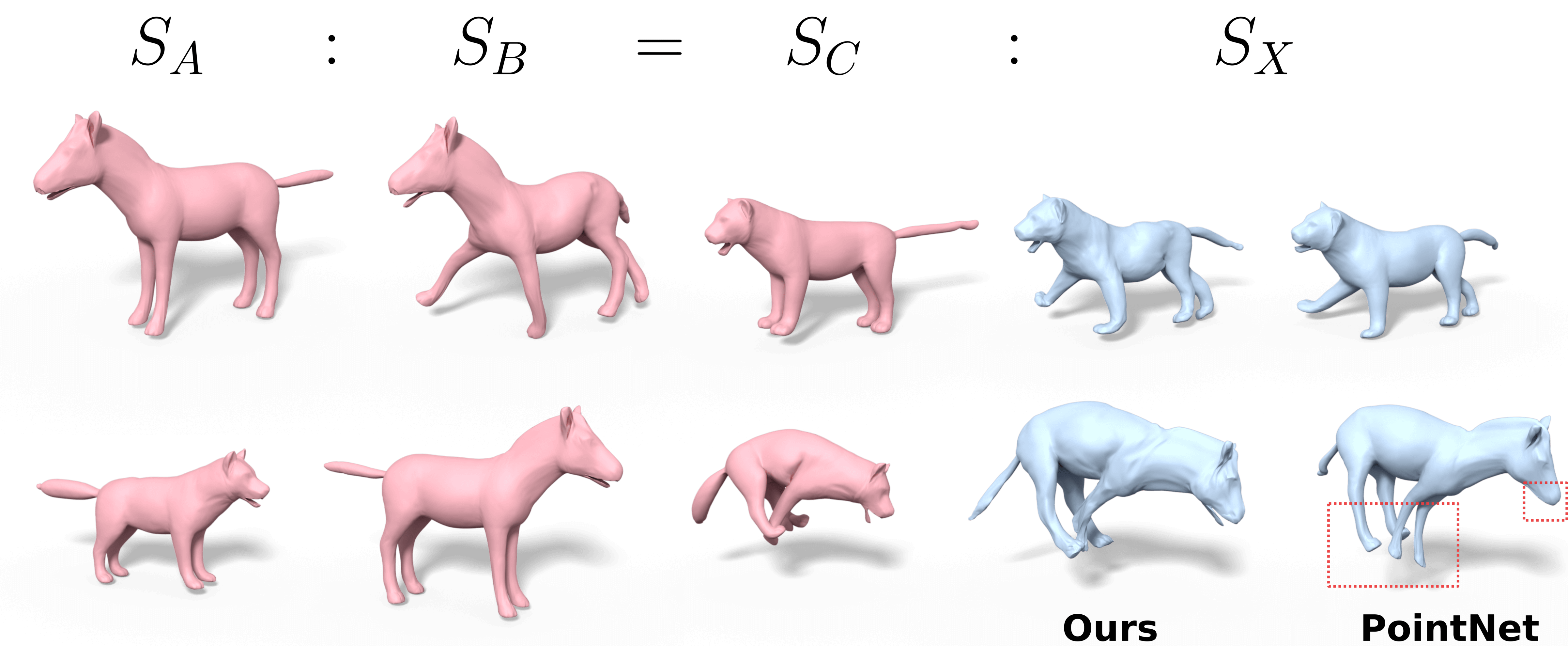}
\caption{\label{fig:animals_analogy} Top row: transferring the pose of $S_B$, from $S_C$ to $S_X$. Bottom row: transferring the animal type of $S_B$, from $S_C$ to $S_X$. PointNet does not maintain the correct pose (bottom row) and does not transfer details such as open mouths correctly.} 
\end{figure}

Lastly, we show analogies among animals in Figure~\ref{fig:animals_analogy}, where we present both pose transfer (top row) and style transfer (bottom row) and comparison to the results of PointNet. 

\section{Shape Interpolation}\label{sec:interp}
\begin{figure*}[!ht]
\centering
\begin{subfigure}{\textwidth}
  \centering
  \includegraphics[width=0.9\linewidth]{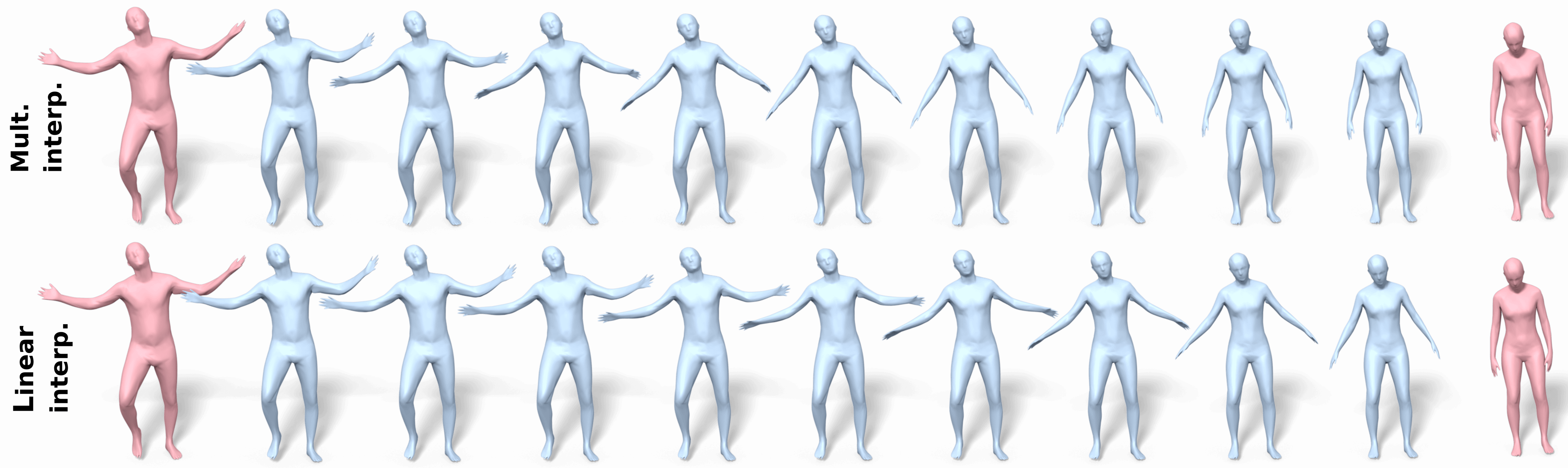}
\end{subfigure}%
\\
\begin{subfigure}{0.55\textwidth}
  \centering
  \includegraphics[width=\linewidth]{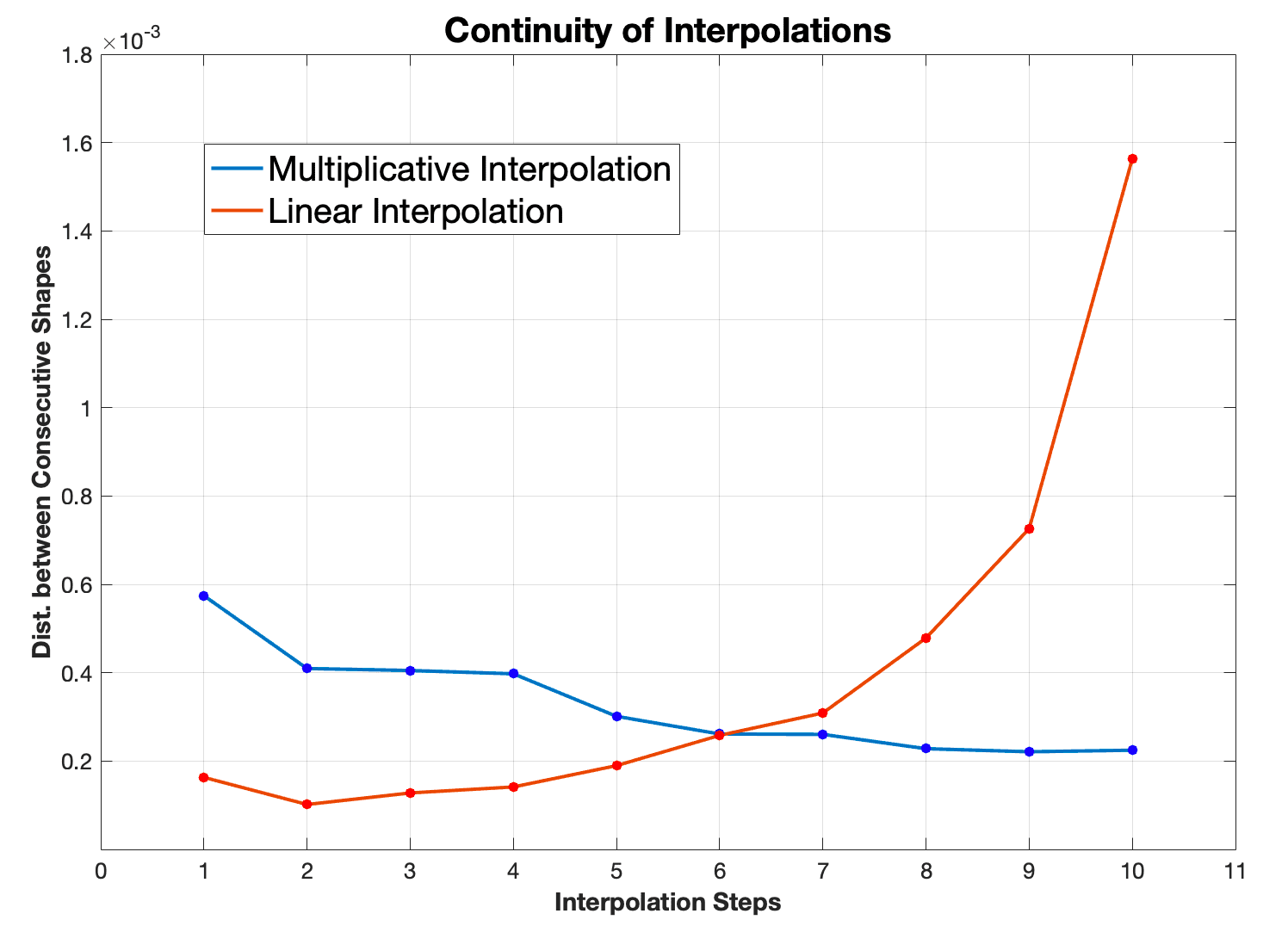}
\end{subfigure}
\caption{\label{fig:lin_vs_geo} Reconstructions regarding shape differences interpolated using multiplicative scheme
  (first row) and using linear scheme (second row). In the bottom panel we plot the distances between consecutive
  reconstructed embeddings for both sequences. The multiplicative scheme clearly delivers more smooth deformation sequence.}
\end{figure*}

\paragraph{Linear Interpolation vs. Multiplicative Interpolation} We note that, since the shape differences are represented by matrices, it is also possible to interpolate shape differences linearly, i.e., $\mathbf{D}(t) = (1-t) \mathbf{D}_0 + t \mathbf{D}_1$. 
However, as we argue in Section~\ref{sec:background}, the \emph{multiplicative} property of shape differences suggests that it is more natural to interpolate the difference operators following Eq.~\eqref{equ:intpolation}. 
To illustrate this point, we show in Figure~\ref{fig:lin_vs_geo} interpolated sequences with respect to the two schemes above -- the multiplicative one in the first row and the linear one in the second row. 
It is visually evident that the former leads to more continuous and evenly deformed sequence. Moreover, we compute the distance between consecutive shapes in both sequences and plot the distributions in the bottom panel of Figure~\ref{fig:lin_vs_geo} as a quantitative verification. 

\paragraph{Baseline Comparison} To make our comparison more complete, we further compare our method to the auto-encoder proposed in 3D-Coded~\cite{groueix2018b}, Multi-Chart GAN proposed in~\cite{ben2018multi}, and a PointNet++~\cite{PNPP} based auto-encoder.

\begin{figure*}[!ht]
  \centering
  \includegraphics[width=0.95\linewidth]{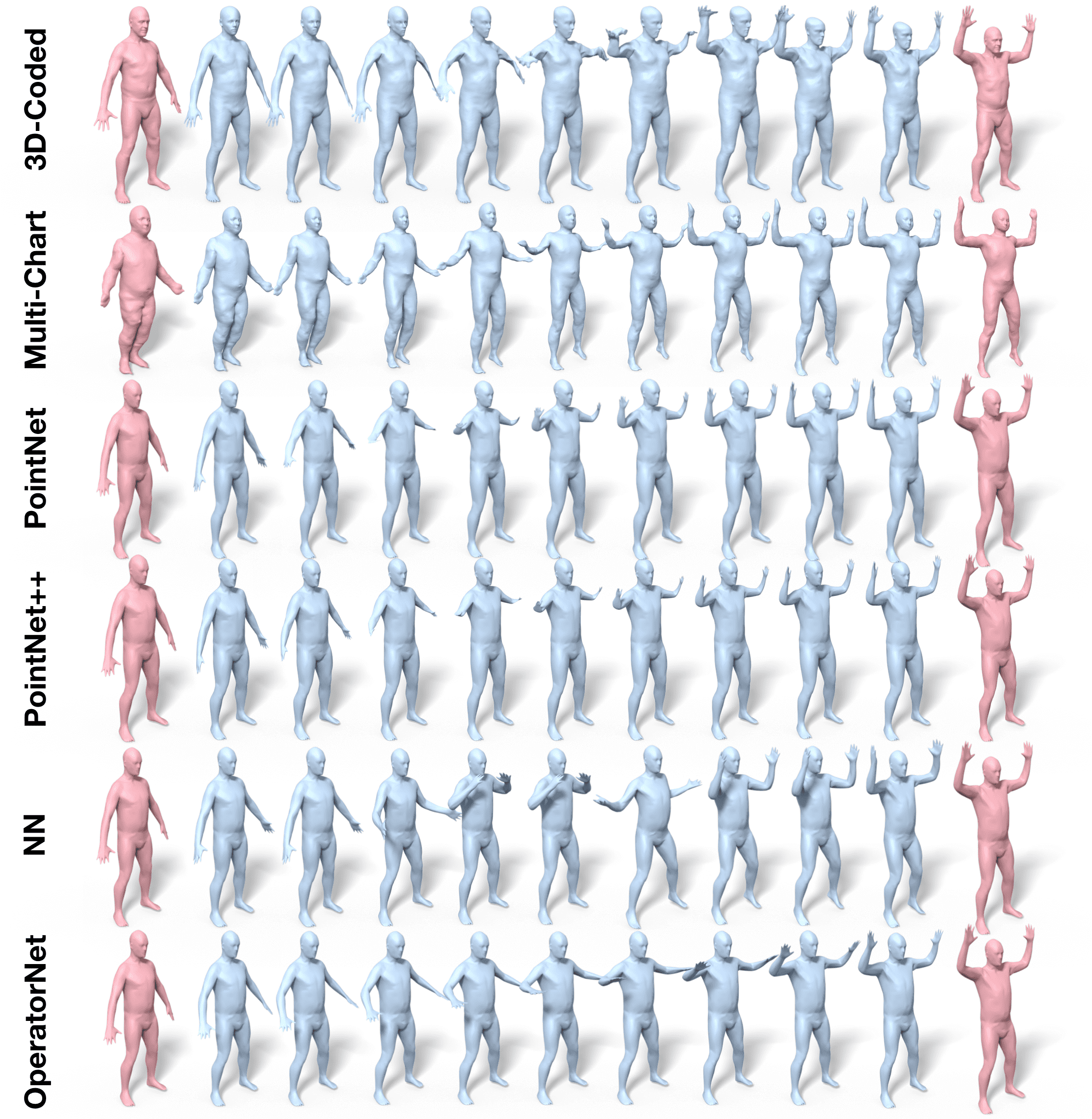}
\caption{\label{fig:comp_pointnet++} From the top row to the bottom row: interpolations via 3D-Coded, Multi-chart GAN, Pointnet, PointNet++ based auto-encoder, Nearest Neighbor in latent space and OperatorNet . } 
\end{figure*}

Regarding 3D-Coded~\cite{groueix2018b} method, we first reconstruct the source and target shapes using their pre-trained model and linearly interpolate the produced latent representations. 
On the other hand, in~\cite{ben2018multi}, a GAN is trained to generate realistic human shapes. In particular, we follow the interpolation scheme described in~\cite{ben2018multi}: first we pick two randomly generated latent vectors $z_1, z_2$, which, via the GAN give arise to two shapes $G(z_1), G(z_2)$. Then, the interpolation between the two shapes is achieved as $G(z(t))$, where $z(t) = (1-t) z_1 + t z_2$.
We randomly generate 1000 shapes using their trained model and pick $G(z_i), i = 1, 2$ that are nearest to the end shapes in the bottom row of Figure~\ref{fig:comp_pointnet++}. 
Lastly, similar to the PointNet baseline, we train an auto-encoder with the PointNet++ encoder and our decoder.

The first row shows the interpolation of~\cite{groueix2018b}. This method generates significant distortions during the interpolation, particularly on the arms. 
In the second row, note that the interpolations from Multi-Chart GAN~\cite{ben2018multi} between the two end shapes are not evenly spaced. For instance, the arms change abruptly during the three middle shapes, while there is little change on that region afterwards. 
As seen in the fourth row of Figure~\ref{fig:comp_pointnet++}, the result of PointNet++ based auto-encoder suffers similar distortions on the arms as that of PointNet (see the third row).
The remaining rows have been shown and analyzed in Figure~\ref{fig:interpolation}.

\end{document}